\documentclass[epj]{svjour}
\usepackage{graphics}
\usepackage{xcolor}
\usepackage{amsmath}
\usepackage{xspace}
\usepackage{amssymb}
\usepackage{cite}
\newcommand{\integral}{\textit{INTEGRAL}\xspace}
\newcommand{\fermilat}{\textit{Fermi}-LAT\xspace}
\newcommand{\fermi}{\textit{Fermi}\xspace}
\begin{document}
\title{Gamma-Ray Astrophysics}
\author{A. De Angelis\inst{1} \and M. Mallamaci\inst{2}
}                     
\offprints{manuela.mallamaci@pd.infn.it}          
\institute{INFN and INAF, Sezione di Padova, Padova, Italy; Universit\`a di Udine; IST and LIP Lisbon \and INFN, Sezione di Padova, Padova, Italy}
\date{Received: date / Revised version: date}
%
\abstract{
High-energy photons are a powerful probe for astrophysics and for fundamental physics in extreme conditions. In recent years, our knowledge of the most violent phenomena in the Universe has impressively progressed thanks to the advent of new detectors for gamma rays, for both ground-based and space-borne observations. This article reviews the present status of high-energy gamma-ray astrophysics, with emphasis on the recent results and a look to the future.
} 
\maketitle
\section{Introduction}\label{intro}
\begin{figure}
\centering    
\resizebox{0.65\textwidth}{!}{
\includegraphics{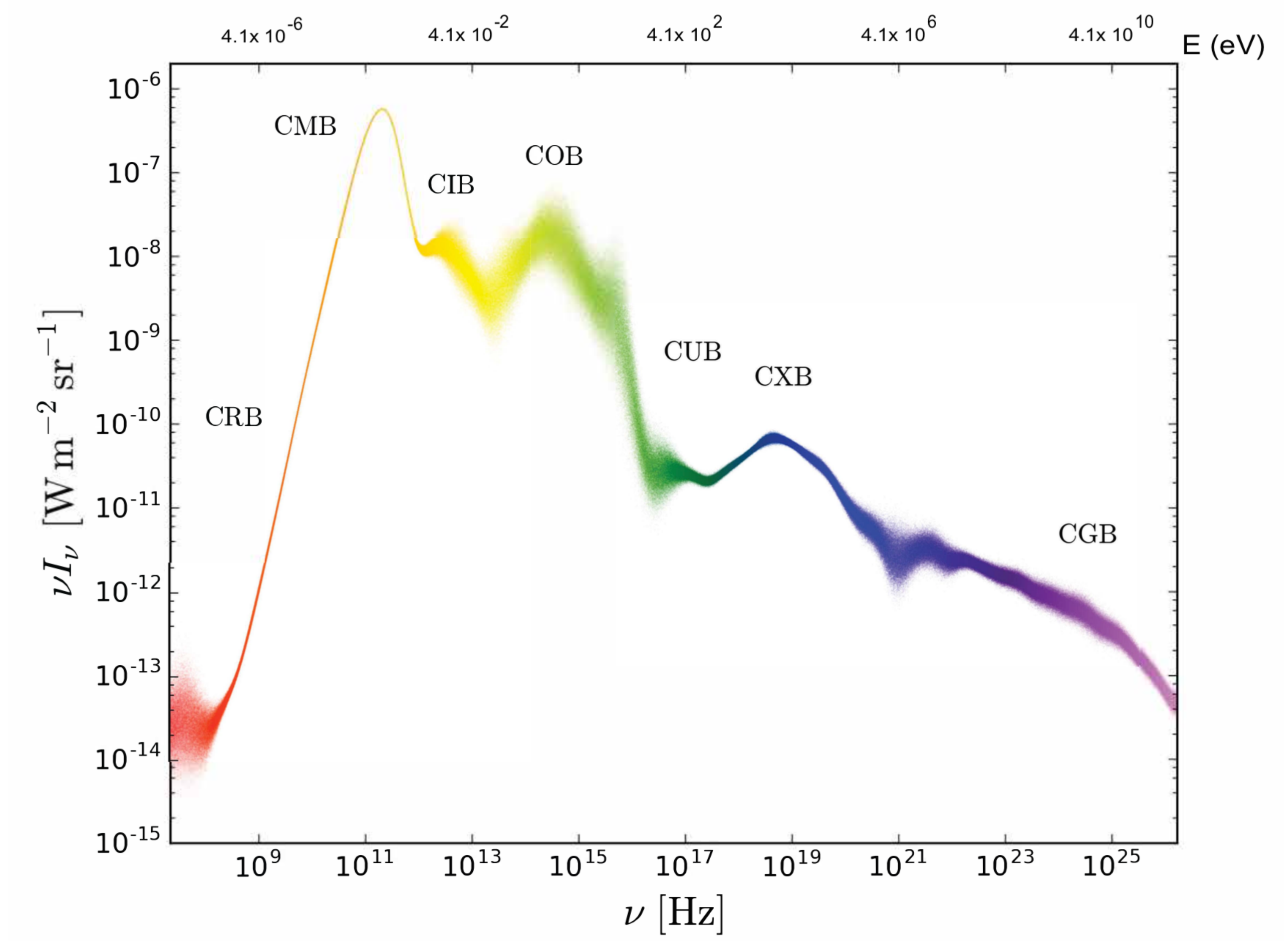}
}
\caption{Energy spectrum of photons measured at different energies, from radio waves to gamma rays. Adapted from \cite{scott}.}
\label{fig:1}       
\end{figure}

Photons are the traditional messengers for astronomical studies and the experimental data collected over  time show us a very complex and fascinating picture of the Universe. The energy spectrum spans  some 20 energy decades, as reported in Fig. \ref{fig:1} \cite{bookap}. In the present review, we will focus on the most energetic photons, indicated as \textit{gamma rays}. In general, the term denotes the electromagnetic radiation above some 100 keV, but we will dedicate the discussion especially to gamma rays in the high energy (between a few MeV and $\sim$ 30 GeV) and very high energy ($\gtrsim$ 30 GeV) regions. There is little doubt on the existence of photons in the PeV-EeV range, but so far cosmic gamma rays have been unambiguously detected in high energy and very high energy domains, the most energetic reaching $\sim$~100~TeV. In recent years, a large number of sources has been detected, revealing the existence of a very  heterogeneous population that lights up the gamma-ray sky, in our Galaxy and beyond. Understanding and modeling the gamma-ray emission are compelling aspects {\em per se}, but have also other facets. Gamma rays are indeed intimately related to all cosmic messengers: cosmic rays, neutrinos and gravitational waves. Besides, they are an indirect probe for questions related to fundamental physics.

The scientific case for gamma-ray studies will be highlighted in Sect.~\ref{sec:science_case}. Some of the main motivations will be addressed, especially in the light of the multi-messenger era we are living now. Sect.~\ref{sec:prod} and~\ref{sec:prop} are dedicated to a brief description of the production and propagation of gamma rays. Observational techniques will be illustrated in Sect.~\ref{sec:instruments}, along with the current generation of instruments, operating on satellites and from ground. Sect.~\ref{sec:sky}, \ref{sec:multi}, \ref{sec:dm} are focused on some of the most exciting results in this field, respectively related to the gamma-ray sky observations, to multi-messenger astronomy and to dark matter searches. We choose to show these results since they  answer to the questions raised in this review. Finally, Sect.~\ref{sec:future} is dedicated to the future of this field. 
\section{The science case}\label{sec:science_case}
The study of gamma rays involves a large field of research, interesting for many reasons. 
First of all, looking at the gamma-ray window of the electromagnetic spectrum is fascinating {\em per se}: it brings us an exceptional view on the non-thermal Universe and it allows us to explore the most violent phenomena happening on very different astronomical scales. In our Galaxy, a number of pulsars, pulsar wind nebulae (PWNe), supernova remnants (SNRs) and micro-quasars have been detected to emit such energetic radiation. Outside the Milky Way, gamma rays have been observed from galaxies with an exceptional rate of star formation and  from ultra-relativistic jets of particles escaping super-massive black holes (SMBHs) located in the center of some galaxies. In addition, extremely energetic transient events called \textit{gamma-ray bursts} (GRBs) populate the gamma-ray sky. Finally, the gamma-ray emission is seen not only from localized sources, but also from diffuse regions in our Galaxy and beyond. The underlying mechanisms of both these localized and diffuse emissions are a subject of study (see Sect.~\ref{sec:sky}).

However,  doing astronomy has nowadays a broader and exciting meaning. The Universe is in fact observed not only through the different windows of the electromagnetic spectrum, but also through other cosmic messengers, i.e. through cosmic rays (CRs), neutrinos and gravitational waves (GWs). In general, gamma rays are the perfect companions for multi-messenger astronomy, as we will highlight in the following.

First of all, gamma-ray production is intimately related to the production of CRs. The latter are charged particles, mainly protons, whose energy spectrum covers a very wide range in energy and flux. Many questions regarding CRs are still open, especially looking at the most energetic ones above $10^{15}$~eV (1 PeV).
The CR spectrum is approximately described by a power law: $\mathrm{dN}/\mathrm{dE} \sim \mathrm{E}^{-\Gamma}$, where $\Gamma$ is  the spectral index. $\Gamma$ is not constant, indicating a change in the properties of CRs, like their acceleration sites and chemical composition.  For energies around $\sim4\times10^{15}$~eV, the flux starts to decrease more steeply: $\Gamma$ changes from about 2.7 to about 3. This feature, marked with the term \textit{knee}, is thought to indicate the maximum acceleration energy of Galactic sources \cite{knee}. The flux exhibits another change in slope around $10^{17}$~eV, where the spectral index becomes about 3.3. This feature is called \textit{second knee} \cite{sknee}. For energies $\sim4.8\times10^{18}$~eV the spectrum flattens and $\Gamma$ returns to about 2.6. The corresponding feature is called \textit{ankle}.  Around $\sim4.2\times10^{19}$~eV  a strong suppression of the flux has been observed~\cite{hires,auger1} and it has reached in recent years a statistical significance of more than 20$\sigma$. This suppression can be ascribed to  energy losses during propagation (the so-called \textit{GZK effect} \cite{Greisen,Zatsepin}) or to an intrinsic limit of sources, that are not able to accelerate particles beyond a certain energy \cite{Hillas}. 
The study of CRs above 1 PeV is extremely challenging for different reasons: on one hand, their flux decreases with increasing energies and the observations can be performed only indirectly with ground-based instruments; on the other hand, CRs suffer magnetic deflection along their path through the Galactic and/or intergalactic medium. Gamma rays instead point to their sources and the Universe is essentially transparent to them up to about 100 GeV (see Sect. \ref{sec:b-up}).  Gamma rays can be therefore used as probes for revealing the sites of CR acceleration, as it will be discussed in Sect.~\ref{sec:CR}.

 According to the bottom-up scenario (i.e. CRs accelerated by astrophysical sources, see Sect. \ref{sec:prod}), gamma rays can be produced by the radiation from charged particles in a magnetic field.
 In addition, both gamma rays and neutrinos can be produced  from the interaction of CRs with nuclear targets, such as molecular clouds. Neutrinos cannot be absorbed nor radiated during their path from the source to the observer, even if they are very difficult to detect. Identifying neutrino sources and their association with gamma ray counterparts  therefore provides unique insights into the long-standing problem of the CR origin. Some interesting results have been obtained and are briefly described in Sect.~\ref{sec:neutrino}.
 
In the light of the recent discoveries~\cite{GW170817}, outlined in Sect.~\ref{sec:GW}, joint GWs and electromagnetic observations have a key role to obtain a more complete knowledge of the sources and their environments, since they provide complementary information. On one hand, GW signals give insights into the physics of the source such as, e.g., the mass and the distance; on the other hand, the presence of  an electromagnetic counterpart  allows us to pinpoint the location of the burst, possibly identifying the host galaxy and properly defining the astrophysical context. Again, we expect that gamma rays are the most suitable electromagnetic counterpart for these kind of studies. 

Finally, the study of gamma rays can address many fundamental physics questions related for example to violation of the Lorentz invariance   or to the existence of exotic particles and objects, like axion-like particles~\cite{darma} and primordial black-holes~\cite{bookap}. In addition, it could elucidate the question of matter-antimatter asymmetry  and, last but not least, aspects related to Dark Matter (DM) particles.  The favorite candidates of DM are the so-called \textit{Weakly Interacting Massive Particles} (WIMPs). They are expected to mutually annihilate, producing gamma rays, electrons, positrons and neutrinos (see Sect.~\ref{sec:t-down}). Therefore gamma rays offer the possibility to probe the presence of DM. Some results about this kind of searches will be described in Sect.~\ref{sec:dm}.
\section{Production of gamma rays}\label{sec:prod}
The spectrum of photons (and neutrinos also) is usually measured as the energy flux F$_\nu$ in erg (or in eV or multiples) per unit area per unit time per unit frequency $\nu$ (in Hz). The spectrum can be characterized by defining its spectral index. Another important quantity is the spectral energy distribution (SED): $\nu$F$_\nu$ in erg cm$^{-2}$ s$^{-1}$, which can be also expressed as:
\begin{equation}
\nu {F}_\nu = {E}^2\frac{{dN}}{dE}
\end{equation}
where $E$ is the energy and $dN/dE$ is the differential energy spectrum.

The main ideas about the origin of gamma rays can be summarised in two classes:
\begin{itemize}
\item{bottom-up models: according to this scenario, the production of gamma rays follows the acceleration of charged particles from  astrophysical sources;}
\item{top-down models: in this case, the observed radiation is the decay product of exotic particles.}
\end{itemize}
 In what follows, some basic concepts about the two scenarios are outlined.
\subsection{Bottom-up scenario}\label{sec:b-up}
In order to interpret the observation of gamma-rays and their relation with other cosmic messengers, it is fundamental to model the electromagnetic emission at the source. Since photons cannot be directly accelerated, a population of charged particles in a certain environment is usually the starting point. The observation of gamma rays is therefore related to questions regarding the astrophysical sources of CRs. In general, the latter are thought to be accelerated through the first-order Fermi mechanism, involving a shockwave moving at relativistic speed~\cite{fermi}. The maximum energy to which a particle can be accelerated, $E_{CR,max}$, can be expressed as: 
\begin{equation}
E_{CR, max}\propto Z \beta \left(\frac{B}{1\, \mu\mathrm{G}} \right)\left(\frac{L}{1\, \mathrm{pc}} \right) \, \mathrm{PeV}\label{Hillas}
\end{equation}
where $Z$ is the charge of the particle,  $L$ is the size of the accelerating region, $B$ the magnetic field permeating it, and $\beta$ is the speed of the shock in unit of light speed~\cite{hillas}. According to this simplified relation (simplified because for example it does not account for radiative losses), there are astrophysical objects with the right combination of size and magnetic field. The latter are therefore able to accelerate particles to ultra-high energies, i.e. above $\sim10^{18}$~eV (1~EeV). 

Typical astrophysical sources in our Galaxy are pulsars, PWNe, SNRs and binary systems. Extragalactic emissions are instead very often associated to the the so-called \textit{Active Galactic Nuclei} (AGN), located in the center of other galaxies.

If the parent population  producing gamma rays is made of leptons or hadrons, we talk about \textit{leptonic} and \textit{hadronic} emission respectively. In particular, the flux of gamma rays of leptonic origin is traced by the electron density and by the radiation fields; the flux of gamma rays of hadronic origin depends on the CR density and the target gas density.

\subsubsection{Leptonic emission}\label{sec:lept}
The emission of gamma rays starting from a population of positron/electrons usually occurs when the astrophysical environment is permeated by a magnetic field and/or when the population of leptons is relativistic. Typical sources thought to be the site for leptonic emission are both Galactic (e.g. pulsars) and extragalactic. 

Relativistic particles accelerated in a magnetic field radiate synchrotron photons. The power loss for a charged particle of mass $M$ and charge $Ze$, given a magnetic field of intensity $B$, can be expressed as:
\begin{equation}
-\frac{dE_{CR}}{dt} \simeq 2.6 \, {\rm \frac{keV}{s}} \left({\frac{Zm_e}{M}}\right)^4  \left({\frac{E_{CR}}{1\,{\rm{keV}}}}\right)^2 \left({\frac{B}{1\,{\rm{\mu G}}}}\right)^2
\end{equation} 
From this equation, it is clear that this kind of process is more important for electrons than for protons. For this reason, the synchrotron is usually ascribed to the acceleration of leptons. Given a population of electrons with an isotropic distribution of pitch angle (i.e. the angle between the velocity and the magnetic field) and energy $E_e$, synchrotron photons will have an energy~\cite{aharonian}:
 \begin{equation}
 E_{sync} = 0.2 \frac{B}{10 \,\textrm{$\mu$G}}\left(\frac{E_e}{1\, \textrm{TeV}}\right)^2 \, \textrm{eV}
 \end{equation}
Therefore, given typical magnetic fields of 10 $\mu$G and a population of electrons with TeV energies, the synchrotron photons will have an energy of about 0.2 eV (i.e. visible light). This mechanism is important also for the emission of gamma rays for reasons that are clarified below.

Another mechanism involving photons and relativistic electrons is the \textit{Inverse Compton} (IC) scattering. In this case, the electrons scatter low energy photons, so that they gain energy at the expense of the kinetic energy of the electrons. This process can occur in two different regimes, depending on the energy of the photon, we call it $E_\gamma$.
\begin{itemize}
\item{the Thomson regime: $E_\gamma \ll m_ec^2$}
\item{the Klein-Nishina regime: $E_\gamma \gg m_ec^2$}.
\end{itemize}
In astrophysical sources, like AGN or the surrounding of SNRs, both synchrotron and IC can take place. In particular, ultra-relativistic electrons with Lorenz factor $\gamma_e\sim$ 10$^{4-5}$ are accelerated in a magnetic field and emit photons up to the infrared/X-ray band. Such photons in turn interact via Compton scattering with their own parent electron population; since electrons are ultra-relativistic, the photon energy can be boosted by a large factor. This process is called \textit{Synchrotron Self Compton} (SSC).\\
Given a population of electrons, whose energy spectrum has a spectral index $\alpha$, and given a blackbody population of soft photons at a temperature T and energy $\epsilon$, we can evaluate the mean energy of gamma rays and the corresponding energy distribution: 
\begin{align}
 E_\gamma \simeq \frac{4}{3} \gamma_e^2 \langle\epsilon\rangle \, ~~~~~~~~~~~~~~~~~~~~~~~~~~\textrm{(Thomson regime)}\\		
 E_\gamma \simeq  \frac{1}{2}  \langle E_e\rangle  \,  ~~~~~~~~~~~~~~~~~~~~~ \textrm{(Klein-Nishina regime)}
 \end{align}
 \begin{align}
 \frac{dN_\gamma}{dE_\gamma} \propto E_\gamma^{-\frac{\alpha+1}{2}} \, ~~~~~~~~~~~~~~~~~~~~~~~~\textrm{(Thomson regime)}\\
  \frac{dN_\gamma}{dE_\gamma} \propto E_\gamma^{-(\alpha+1)}\ln E_\gamma \,  ~~~~~~~~~ \textrm{(Klein-Nishina regime)}
 \end{align}
 A  useful expression relating the electron and the photon energy is~\cite{bookap}:
\begin{equation}
E_\gamma \simeq 6.5 \left(\frac{E_e}{1\, \mathrm{TeV}}\right)^2 \left(\frac{\epsilon}{\mathrm{meV}}\right) \mathrm{GeV} \, ,
\end{equation}
The Compton component can peak at GeV-TeV energies.
\subsubsection{Hadronic emission}\label{sec:had}
Gamma rays can be the product of the interaction between accelerated protons (and/or heavier nuclei) and the astrophysical environment. 
In general, a hadron can collide with a target of nucleons (for example, a molecular cloud), initiating a hadronic cascade.  Almost the same number of $\pi^+$, $\pi^-$, $\pi^0$ are produced, due to isospin simmetry. Given to the short lifetime, $\pi^0$ immediately decay into two gamma rays, having approximately half the energy of $\pi^0$. This is referred to as \textit{hadron-nucleon collision}, or simply \textit{pp} interaction in the simple case of a collision between protons. Detailed Monte Carlo simulations are required in general to describe the shower development, according to models like the recent versions of QGSJet~\cite{qgs}, EPOS~\cite{epos}, SYBILL~\cite {sybill}.

Gamma rays can originate also from the interaction between protons and a sea of photons, coming for example from synchrotron radiation or  from bremsstrahlung of accelerated electrons.
This kind of interaction is called \textit{photoproduction}. It has a small cross section (fraction of mb) and it is therefore important in environments where the target photon density is much higher than the matter density. 
In particular, photoproduction occurs mainly via the $\Delta^+$ resonance:
\[
p+ \gamma \rightarrow \Delta^+ \rightarrow \pi^++n 
\]
\[
p+ \gamma \rightarrow \Delta^+ \rightarrow\pi^0+p 
\]
The cross sections of these two processes at the $\Delta$ resonance are approximately in the ratio 1:2, due to isospin balance. Again a $\pi^0$ decays into a pair of photons. Their energy will depend on the average momentum fraction carried by the secondary pions relative to the primary particle and on the average fraction of the pion energy carried by the photon. Approximately we can state that $E_{\gamma}\sim E_p/10$~\cite{bookap}.

A characteristic of hadro-production of gamma rays is a peak at $\simeq m_\pi c^2/2  \simeq 67.5$ MeV  in the spectral energy distribution, which can be related to a component from $\pi^0$ decay; this feature, which is almost independent of the energy distribution of $\pi^0$ mesons and consequently of the parent protons, is called the \textit{``pion bump''},
and can be explained as follows. In the rest frame of the neutral pion, both photons
have energy $E_\gamma \simeq 67.5$~MeV and momentum opposite to each other. Once boosted for the energy $E$ of the
emitting $\pi^0$, the probability to emit a photon of energy $E_\gamma$ is constant over the range of kinematically allowed energies  (the interval between $E(1-v/c)/2$ and $E(1+v/c)/2$). The spectrum of gamma rays for an arbitrary distribution of neutral pions is thus a superposition of rectangles for which only one point at $m_\pi c^2/2$ is always
present. This should result in a spectral maximum independent of the energy distribution of parent pions. The ``pion bump'' is considered as a spectral signature of hadronic processes at work (see Sect. \ref{sec:CR}).

It is worth mentioning that gamma rays production from hadrons is always accompanied by the production of neutrinos. If the $\pi^0$ decays immediately into two gamma rays, the charged pions decay as  $\pi^+ \rightarrow \mu^+\nu_{\mu} $ and $\pi^- \rightarrow \mu^-\bar{\nu}_{\mu}$, with muons decaying into $\mu^+ \rightarrow e^+\nu_e\bar{\nu}_{\mu}$ and $\mu^- \rightarrow e^-\bar{\nu}_e\nu_{\mu}$. 
An approximate relation holds at emission between the spectral production rates of neutrinos and gamma rays~\cite{bookap}:
\[
E_\nu^2 \frac{dN_\nu(E_\nu)}{dE_\nu} \sim \frac{3K}{4}  E^2_\gamma \frac{dN_\gamma(E_\gamma)}{dE_\gamma} \]
with $K = 1/2$ (or =2) for photo-hadronic (or hadronuclear) processes. Depending on the source optical depth, gamma rays may escape or further cascade, complicating time and energy correlation between neutrinos and electromagnetic counterparts.
A simultaneous detection of both gamma rays and neutrinos would be however an important signature of hadro-production (see Sect. \ref{sec:neutrino}), and therefore of CR acceleration.
\subsection{Top-down scenario: dark matter}\label{sec:t-down}
According to a top-down scenario, gamma rays might be produced from exotic particles, in particular from their decay or their annihilation. In particular, gamma rays are believed to possibly originate from DM. Among the most popular DM candidates, there are the weakly interacting massive particles (WIMPs), with masses and coupling strengths at the electroweak scale. The expected flux of gamma rays from annihilation processes can be expressed as:
\begin{equation}
\phi_\gamma=\frac{1}{4\pi}\,\underbrace{\frac{\langle
\sigma_{ann}
v\rangle}{2m^2_{DM}}\,\frac{dN_{\gamma}}{dE}}_{{\rm{Particle\,
Physics}}}\,\underbrace{\int_{\Delta\Omega-l.o.s.} dl(\Omega) \rho^2_{DM}}_{\rm{Astrophysics}} \, .\label{eq:dmrate}
\end{equation}
where the first factor depends on particle physics properties of DM: $\langle\sigma_{ann}v\rangle$ is the velocity-weighted annihilation cross section (also called \textit{annihilation rate}) of DM particles of mass $m_{DM}$ and $dN_{\gamma}/dE$ is the gamma-ray spectrum per annihilation event. The second factor is the line-of-sight integral of the squared DM density and it is therefore related to its astrophysical distribution. It is called \textit{boost factor}. DM-induced gamma rays could exhibit sharp spectral signatures, like for instance $\gamma\gamma$ or $Z\gamma$ annihilation lines, with energies strictly related to the WIMP mass. However, since  WIMPs are electrically neutral, these processes are loop suppressed and therefore should be rare. WIMP-induced gamma rays are thus expected to be dominated by a relatively featureless continuum of by-products of cascades and decays (mostly from $\pi^0$) following the annihilation in pairs of quarks  or leptons. As mentioned, and as indicated in Eq.~\ref{eq:dmrate}, the flux of resulting gamma rays depends quadratically on the DM density along the line of sight of the observer. This motivates searches on targets, where one expects DM density enhancements. Among these targets, we find the Galactic center (GC), galaxy clusters, and nearby dwarf spheroidal galaxies, as will be discussed in Sect. \ref{sec:dm}. 
\section{Propagation of gamma rays}\label{sec:prop}
\begin{figure}
\centering    
\resizebox{0.45\textwidth}{!}{
\includegraphics{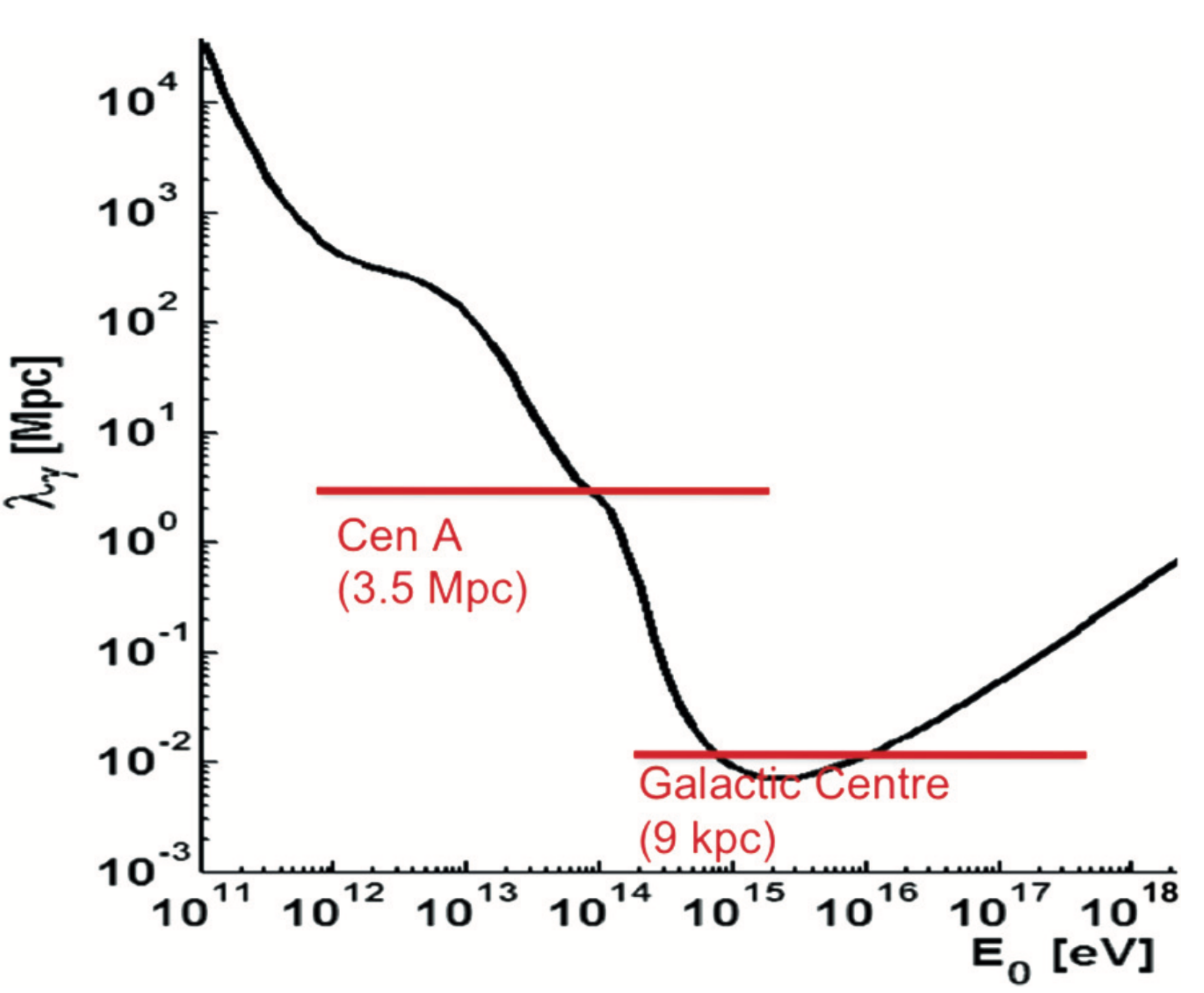}
}
\caption{Mean free path of gamma rays of energies between 100 GeV and 1 EeV, at $z=0$~\cite{deangelis2013}. See text for details.}
\label{fig:horizon}       
\end{figure}
After being produced, gamma rays propagate through the intergalactic/galactic space. Although they are not deflected by magnetic fields, they can interact with photon fields. The photon background is shown in Fig.~\ref{fig:1}. The maximum density corresponds to the so-called \textit{Cosmic Microwave Background} (CMB) with $\sim$ 410 photons/cm$^3$ and average energy of  $\sim$ 0.6 meV. This is the residual electromagnetic energy from the Big Bang. Another photon background is the so-called  \textit{Extragalactic Background Light} (EBL). This radiation was mostly emitted during star formation; its spectrum exhibits one peak at about 8 meV (far IR) and one at about 1 eV (near IR, with a tail extending in the visible and UV).

The dominant process for the absorption of gamma rays is pair creation:
\[
 \gamma + \gamma_{{\rm{background}}} \rightarrow{} e^+ + e^-  \, .
\]
Given $\epsilon$, the energy of the target (background) photon, the process is kinematically allowed for:
\begin{equation}\label{eq.sez.urto01012011}
\epsilon > {\epsilon}_{\rm thr}(E,\varphi) \equiv \frac{2 \, m_e^2 \, c^4}{ E \left(1-\cos \varphi \right)}
\end{equation}
where $\varphi$ denotes the scattering angle, $m_e$ is the electron mass, $E$ is the energy of the incident photon.
The corresponding cross section~\cite{breitwheeler} is:
\begin{equation}\label{eq:bhe}
\sigma_{\gamma \gamma}(E,\epsilon,\varphi) \simeq 1.25 \cdot 10^{-25} \,  W(\beta) \,  {\rm cm}^2
\end{equation}
with $W(\beta)$ a term depending on E, $\epsilon$ and $\varphi$ through the electron speed $\beta$, expressed in natural units, in the center-of-mass.
According to~\cite{breitwheeler}, for an isotropic background of photons, the cross section is maximized for:
\begin{equation}\label{eq.sez.urtosps1}
\epsilon (E) \simeq \left(\frac{900 \, {\rm GeV}}{E} \right) \, {\rm eV} \, .
\end{equation}
This means that the EBL plays the leading role in the absorption for gamma rays of energies $10 \, {\rm GeV} \leq E < 10^5 \, {\rm GeV}$.  For $10^5 \, {\rm GeV} \leq E < 10^{10} \, {\rm GeV}$ the interaction with the CMB becomes dominant. At energies $E \geq 10^{10} \, {\rm GeV}$ the main source of opacity of the Universe is the radio background~\cite{Protheroe}.

Neglecting the expansion of the Universe and starting from the cross section (Eq.~\ref{eq.sez.urtosps1}), the mean free path $\lambda$ can be evaluated and is shown in Fig.~\ref{fig:horizon}~\cite{deangelis2013}. The interaction with photon backgrounds causes a horizon beyond which gamma rays are strongly absorbed. This horizon gets closer as the energy increases. At PeV energies, the horizon is close to the distance of our own GC. 
\section{The instruments}\label{sec:instruments}
The atmosphere is opaque to high energy photons beyond the optical waveband. For this reason, high-energy astrophysics required the advent of space-based experiments. In particular, it began with the discovery of the first cosmic X-ray source by Giacconi and Rossi in 1962 \cite{Xastro}.

\begin{figure}
\centering    
\resizebox{0.60\textwidth}{!}{
\includegraphics{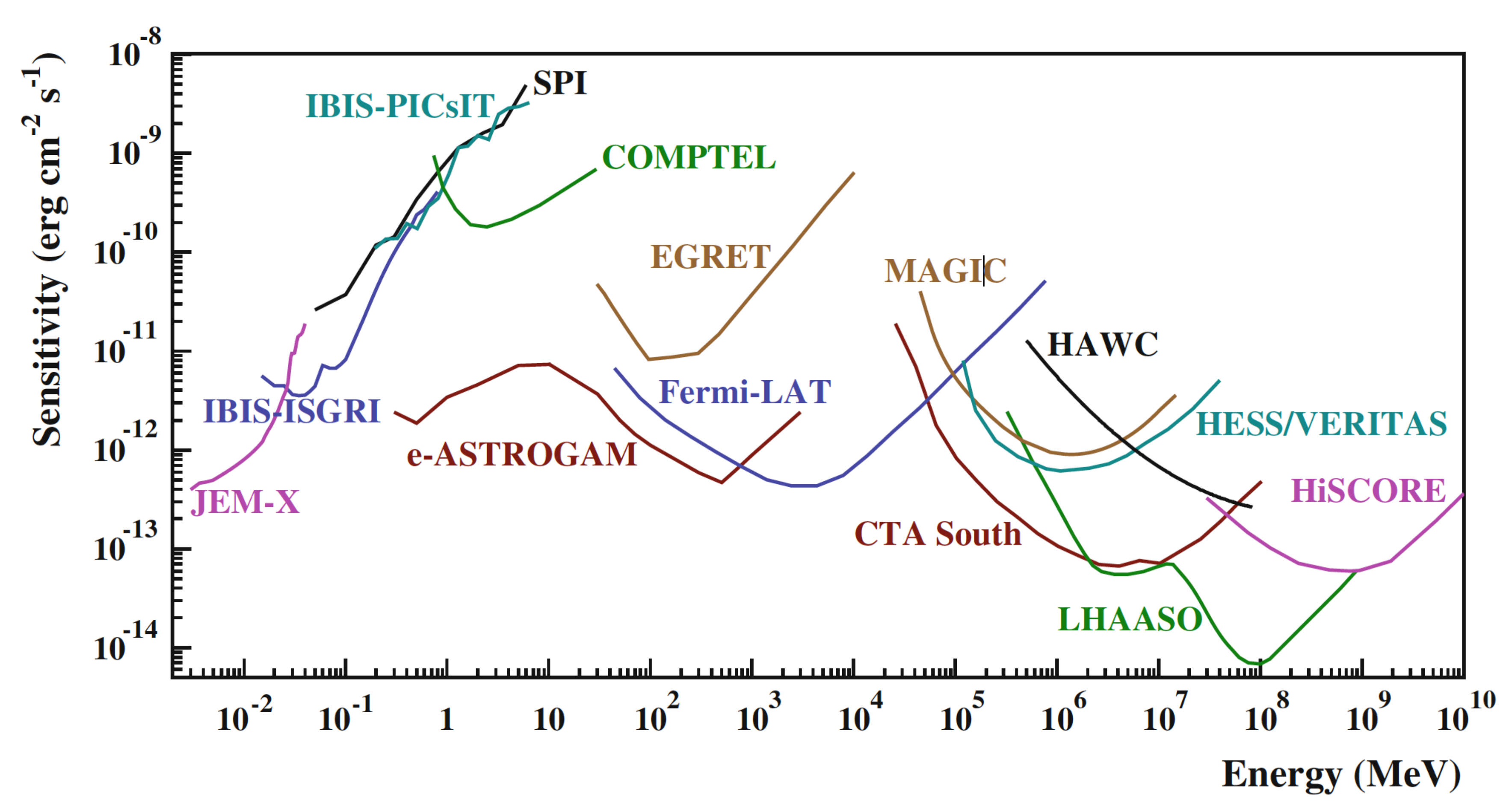}
}
\caption{Point source continuum differential sensitivity of different X- and gamma-ray instruments. The curves for \integral/JEM-X, IBIS (ISGRI and PICsIT), and SPI are for an effective observation time $T_{\rm obs}$ = 1 Ms. The COMPTEL and EGRET sensitivities are given for {the typical observation time accumulated during the $\sim$9 years of the {\it CGRO} mission. The  sensitivity is for a high Galactic latitude source in 10 years of  observation in survey mode}. For MAGIC, VERITAS, H.E.S.S., and CTA, the sensitivities are given for $T_{\rm obs}$ = 50 hours. For HAWC $T_{\rm obs}$ = 5 yr, for LHAASO $T_{\rm obs}$ = 1~yr, and for HiSCORE $T_{\rm obs}$ = 1000 h. This figure shows also the sensitivity of e-ASTROGAM (see Sect. \ref{sec:future}), calculated at $3\sigma$ for an effective exposure of 1 year and for a source at high Galactic latitude.}
\label{sensitivity}       
\end{figure}

Two kinds of gamma-ray instruments exist:  space-based and ground-based detectors. These two typologies are complementary. The experimental spectrum of gamma rays spans indeed 7 decades in energy and about 14 in flux, rapidly decreasing towards high energies. It is therefore clear that the larger is the energy, the larger should be the effective area, defined as the product of the geometrical area and the detector efficiency. 
Because of the cost of space technology, the geometrical area cannot however exceed $\sim$ 1 m$^2$. This aspect makes space-based detectors more appropriate for measuring gamma rays in the MeV -- mid-GeV energy range.
Going  to higher energies, large detection areas are needed and can be deployed only at ground, exploiting the fact that, for energies above $\sim$ 30 GeV, the so-called \textit{electromagnetic air showers} start to become detectable (whereas if the energy is too low, the shower cannot develop properly).  When a gamma ray enters  the
atmosphere, it generates a cascade of secondary particles: the photon converts into pairs of $e^+e^-$ at high altitude and each high-energy $e^\pm$ radiates
secondary gamma rays mostly  through bremsstrahlung, which further convert into $e^+e^-$ pairs of lower energies. \\
In the following, both space- and ground-based techniques are discussed, focusing on some historical remarks~\cite{bernard} and briefly describing the past and current generations of gamma-ray detectors with their key characteristics, such as the field of view (FoV), the duty cycle, the background (mainly CRs) rejection, the angular and  energy resolution, the sensitivity. The main figures of merit of the current detectors are reported in Table \ref{comparison}. Fig.~\ref{sensitivity} shows  the sensitivity for past and current gamma-ray detectors, along with future ground-based experiments like CTA, LHAASO and HiSCORE and a possible future space mission, e-ASTROGAM. The future directions are discussed in Sect.~\ref{sec:future}.
\begin{table}
\begin{center}
\begin{tabular}{|c|c|c|c|}
\hline
\hline
{Quantity} & $Fermi$  & {IACTs} & {EAS}\\
\hline
Energy range & 20\,MeV${{-}}$200\,GeV & 100\,GeV${{-}}$50\,TeV  & 400\, GeV${{-}}$100\,TeV \\
Energy res.     & 5--10\,\%            & 15--20\,\% & $\sim$ 50\,\%\\
Duty cycle            & 80\,\%              & 15\,\%  & $>$ 90\,\%\\
FoV & $4 \pi / 5$       & 5 deg $\times$ 5 deg & $4 \pi / 6$\\
PSF (deg)      & 0.1            & 0.07              & 0.5 \\
Sensitivity  & 1\,\% Crab (1\,GeV)  & 1\,\% Crab (0.5\,TeV)   & 0.5 Crab (5\,TeV) \\
\hline
\end{tabular}
\end{center}
\caption{A comparison of the characteristics of Fermi, the IACTs and of the Extensive Air Showers (EAS) particle detector arrays. Sensitivity computed over one year for Fermi and the EAS, and over 50h for the IACTs.}
\label{comparison}    
\end{table}

\subsection{Space-based detectors}\label{sec:space}
Space-based telescopes can measure gamma rays between $\sim$ 300 keV and $\sim$ 300 GeV, limited by flux. As compared to soft X-ray astronomy, space-based gamma-ray astronomy faces additional challenges. One of them is that gamma rays above some MeV cannot be focused and have to be detected through their interaction products.  As a consequence,   to measure the direction of the incoming gamma rays and their energy, two sub-detectors are needed: a tracker and a calorimeter. In addition, an anti-coincidence detector (ACD) is fundamental for rejecting the background due to CRs. 

Two are the mechanisms through which cosmic photons can interact in the detector: Compton scattering and pair creation, the transition energy being around $E_\gamma$ $\sim$ 10-20 MeV. 
The tracker is therefore made of some active material that tracks  interaction products, and, if needed, some passive material enhancing the interaction probability. The effective area depends on the fraction of converted gamma rays, which increases with the amount of material. Below $\sim$~10 GeV, the angular resolution is limited by multiple scattering, which also increases with the mass of the converter. A compromise must be therefore found between increasing the effective area and degrading the angular resolution. In general, the effective area cannot exceed $\sim$ 1 m$^2$, since these detectors have a large cost, dominated by that of launch and by the strong requirements for the instruments which must be sent into space. 

 Shortly after the birth of X-ray astronomy, the first attempts to detect cosmic gamma rays with balloon-borne detectors were unsuccessful due to the high level
of secondary gamma rays produced by CRs in the atmosphere.  The first gamma-ray satellites were launched between the end of the 1960s and the beginning of the 1970s:
the OSO-3 (Orbiting Solar System) satellite in 1967-1968 provided the first  evidence that the Milky Way was a bright source of gamma rays above 50~MeV  \cite{oso-3};  SAS-2 (1972-1973; $E > 35$~MeV)   \cite{sas-2} 
 revealed the diffuse emission of the Galaxy and discovered the Crab and Vela nebulae and the periodic signals from their pulsars;  COS-B (1975-1982; $E>100$~MeV) produced a catalog of 25
sources, all Galactic except for one, the quasar 3C 273 \cite{cosb}.

Later on, in the 1980s, the Compton Gamma-ray Observatory (\textit{CGRO}) was launched, taking data from 1991 until 2000. It comprised four instruments that covered six decades of the electromagnetic spectrum, from 30 keV to 30 GeV: the Burst And Transient Source Experiment (BATSE), the Oriented Scintillation Spectrometer Experiment (OSSE), the Imaging Compton Telescope (COMPTEL), and the Energetic Gamma Ray Experiment Telescope (EGRET). COMPTEL \cite{COMPTEL} used the Compton effect for reconstructing an image of a gamma-ray source in the energy range from 1 to 30 MeV, by using a liquid scintillator as passive material and NaI crystals as active material absorbing the scattered photon. EGRET \cite{EGRET} had a tracker made of spark chamber modules. Its calorimeter was a block of NaI, without segmentation, and the ACD was a monolithic dome-shaped scintillator. Between the tracker and the calorimeter, there was a time of flight coincidence system. 
The Third EGRET catalog  
\cite{egret_3cat} revealed 271 sources, among which were many AGN, inaugurating the field of extragalactic
gamma-ray astronomy at high energies.

A new era for  space-based observations started in June 2008, when the \fermi gamma-ray telescope (originary called GLAST) was launched.  It is currently operating and it is the largest gamma-ray space-based detector ever built up to now. It uses particle physics technology and was preceded by the all-Italian pathfinder   AGILE (Astro-rivelatore Gamma a
Immagini LEggero), launched in 2007 \cite{AGILE}. \fermi \cite{fermitel} is composed by the spacecraft and two instruments: the Large Area Telescope (LAT)   and the \fermi Gamma Burst Monitor (GBM). The structure of the LAT consists mainly of a tracker, an ACD  and a calorimeter. Its energy range goes from 20 MeV to about 300 GeV and above, thanks to the calorimeter and ACD. The GBM  consists of two types of detectors: the NaI (8-900 keV) and BGO (250 keV - 40 MeV) detectors, and complements the LAT observations of transient sources. 

Finally DAMPE \cite{DAMPE}, launched in 2015, has a structure and an effective area similar to AGILE. It is however characterized by an imaging calorimeter of about 31 radiation lengths thickness, made up of 14 layers of Bismuth Germanium Oxide (BGO) bars in a hodoscopic arrangement - this is the deepest calorimeter ever used in space.

For lower energies, going from 100 keV to few MeV, Swift \cite{Swift} and \integral \cite{INTEGRAL}, in particular with its instrument JEM-X \cite{JEM-X}, are monitoring the sky. COMPTEL and INTEGRAL detected
gamma-ray lines presumably from electron-positron annihilation (0.511 MeV), as well as from the radioactive decay of nuclei like $^{26}$Al, $^{44}$Ti and $^{60}$Fe (1-2 MeV region), which trace the nucleosynthesis activity of supernovae \cite{diehl}. In the future the proposed MeV-GeV missions e-ASTROGAM and AMEGO could dramatically improve these results \cite{wb}.


\subsection{Ground-based detectors}\label{sec:ground}
One of the main differences between space-based and ground-based detectors is that the latter are able to observe gamma rays in the very high energy domain, i.e. between GeV and TeV energies. In particular, above some GeV, a gamma ray entering  the atmosphere generates an electromagnetic cascade that can be detected:
\begin{itemize}
\item Either by detecting the Cherenkov radiation of charged particles in air (Cherenckov technique). Most   secondary charged particles have a speed larger than light in air and emit therefore Cherenkov radiation, which can be detected by special telescopes. The effective detection area is comparable to that of the light pool on ground, i.e. a few $10^4$~m$^2$.
\item Or by directly detecting the charged particles reaching ground (extensive air shower, or EAS, technique).
\end{itemize}

\subsubsection{The Imaging Air Cherenkov Technique}
The Cherenkov technique is currently providing the best results.
Cherenkov light emission induced by CRs in the atmosphere was detected for the first time in 1952 by W. Galbraith and J.V. Jelley \cite{Galbraith 1953}. This led A. Chudakov to start the first systematic studies in this field   \cite{Chudakov 1958}. In 1959, at the International Cosmic Ray Conference (ICRC) in Moscow, G. Cocconi \cite{Cocconi 1960} suggested to measure sources of gamma rays at TeV energies, and predicted that the Crab Nebula was a strong source of TeV gamma rays and that a detector with a large angular resolution could   reject the isotropic CR background.
This idea stimulated further work to use Cherenkov radiation, as suggested by G.T. Zatsepin and A. Chudakov \cite{Zatsepin 1961}. In the early 1960s, the first Atmospheric Cherenkov Telescope (ACT) designed for gamma-ray observations was built in Crimea.
A landmark was set by the construction of the 10~m diameter ACT Whipple, completed in 1968 on Mount Hopkins  in Arizona. Later on, the 37 pixel imaging camera proposed by T. Weekes in 1981 \cite{Weekes 1981} was completed in 1983, making it a IACT (Imaging ACT).
The instrument allowed an  imaging analysis known as the \textit{Hillas parameters method},  proposed by A.M. Hillas in 1985 \cite{Hillas 1985}, that greatly improved  background rejection. The intensity and area of the image provide an estimate of the shower energy, while the image orientation is related to the shower direction. The shape of the image is characteristic of the nature of the events and is used to reject the background from charged particles.
This was decisive for obtaining the first successful localization of the first gamma ray emitter: the Crab Nebula was detected above 0.7 TeV  in 1989 by the Whipple collaboration \cite{Weekes 1989}, 37~years after the initial Cherenkov light pulse observation.

Given a primary photon of 100 GeV, about 10 Cherenkov photons per square meter  arrive at the level of mountain altitudes around 2000 m above sea level (a.s.l.). A collection area of 100 m$^2$ is therefore sufficient to detect gamma ray showers. Since the Cherenkov light is faint, clear and almost dark nights are required for observations. As a consequence, this kind of instruments are characterized by a low duty cycle of  about 15\%. In addition, they have a small  FoV ( $\lesssim 5^\circ$), but a high sensitivity and a low energy threshold.

HEGRA \cite{HEGRA} on the Canary Islands and CANGAROO \cite{CANGAROO} in Australia were the second generation of IACTs.  
The current generation of instruments is mostly composed by H.E.S.S. in Namibia \cite{HESS}, MAGIC in the Canary Islands \cite{MAGIC} and VERITAS in Arizona
\cite{VERITAS}.
VERITAS (Very Energetic Radiation Imaging Telescope Array System) is an array of four IACTs located at the Whipple Observatory   at an altitude of 1300 m a.s.l. Starting operation in 2007, VERITAS observes gamma rays in the range between $\sim$100 GeV and $\sim$30 TeV. 
MAGIC (Major Atmospheric Gamma-ray Imaging Cherenkov Telescopes) consists of two IACTs with reflectors with a diameter of 17 m, located on the Roque de los Muchachos Observatory (ORM), on the Canary Island of La Palma,   at an altitude of 2200 m a.s.l.. It started operation in 2004 with a single telescope; in  2009, a second telescope was added, improving  sensitivity and angular resolution. The energy threshold is as low as 30 GeV. 
H.E.S.S. (High Energy Stereoscopic System) is an array of five IACTs located in   Namibia at an altitude of 1800 m a.s.l.. H.E.S.S. started operation in 2004 with an array of four IACTs, each with a 13 m diameter reflector. In 2012, a fifth telescope with a large 28 m reflector was added in the center of the array. Thanks to its large mirror area, it can be triggered by gamma rays with energies as low as 30 GeV. 
In addition, FACT (First G-APD Cherenkov Telescope) is a single IACT \cite{FACT} located at the ORM.   It started  observations in 2011. With respect to the other IACTs, FACT can continuously operate during bright moonlight, because it exploits the Silicon PhotoMultiplier (SiPM) technology for its camera. 

\subsubsection{The EAS Technique}
This technique observes the secondary particles from gamma-ray-induced air showers reaching ground. EAS detectors have a high duty cycle and a large FoV, but relatively poor sensitivity. The energy threshold   is rather large -- a shower initiated by a 1 TeV photon typically has its maximum at about 8 km  a.s.l.. The principle of operation is the same used for the detection of CRs above 1 PeV.  {The density of the secondary particles and their arrival times allow the reconstruction of the shower geometry; gamma-hadron discrimination can be achieved through the geometry and possibly the muon content.}

Past examples of ground-based instruments were MILAGRO \cite{MILAGRO}, Tibet AS-gamma \cite{Tibet-AS} and ARGO-YBJ \cite{ARGO}. MILAGRO was a water-Cherenkov instrument located in New Mexico (at an altitude of about 2600 m a.s.l.). It exploited the Cherenkov light produced by the secondary particles of the shower entering the water pool instrumented with photomultipliers. Tibet AS-gamma (located at 4100 m a.s.l.) used a sparse array of scintillator-based detectors.  ARGO-YBJ, located close to Tibet AS-gamma, was   an array of resistive plate counters (RPCs). This approach ensured efficient collection and hence a lower energy threshold. 

Currently,  the HAWC (High Altitude Water Cherenkov) detector is in operation \cite{HAWC}. It is an array of 300 water-Cherenkov detectors, covering an instrumented area of about 22000 m$^2$. Each detector consists of a tank  of 7.3 m diameter  4.5 m deep, filled with purified water and containing three PMTs of 20 cm diameter. Photons traveling through the water produce electrons and positrons, resulting in Cherenkov light emission.  The HAWC observatory is located at an altitude of 4100 m a.s.l. in Mexico. The construction and installation of the full array was completed in 2015. HAWC monitors 2/3 of sky with an instantaneous FoV of $\sim$ 2 sr and with a duty cycle of 90\%. The one-year survey sensitivity of HAWC is $\sim$5-10\% of the flux of the Crab Nebula.
\section{The Universe through gamma rays}\label{sec:sky}
Our knowledge of  gamma rays has enormously improved over the last decade. In particular, a look at the gamma-ray sky shows us the presence of (1) a diffuse background of Galactic origin, (2) a faint diffuse emission of extragalactic origin and (3) a very heterogeneous population of localized emitters. Each of these components is briefly discussed.

\begin{figure}
\centering    
\resizebox{0.45\textwidth}{!}{
\includegraphics{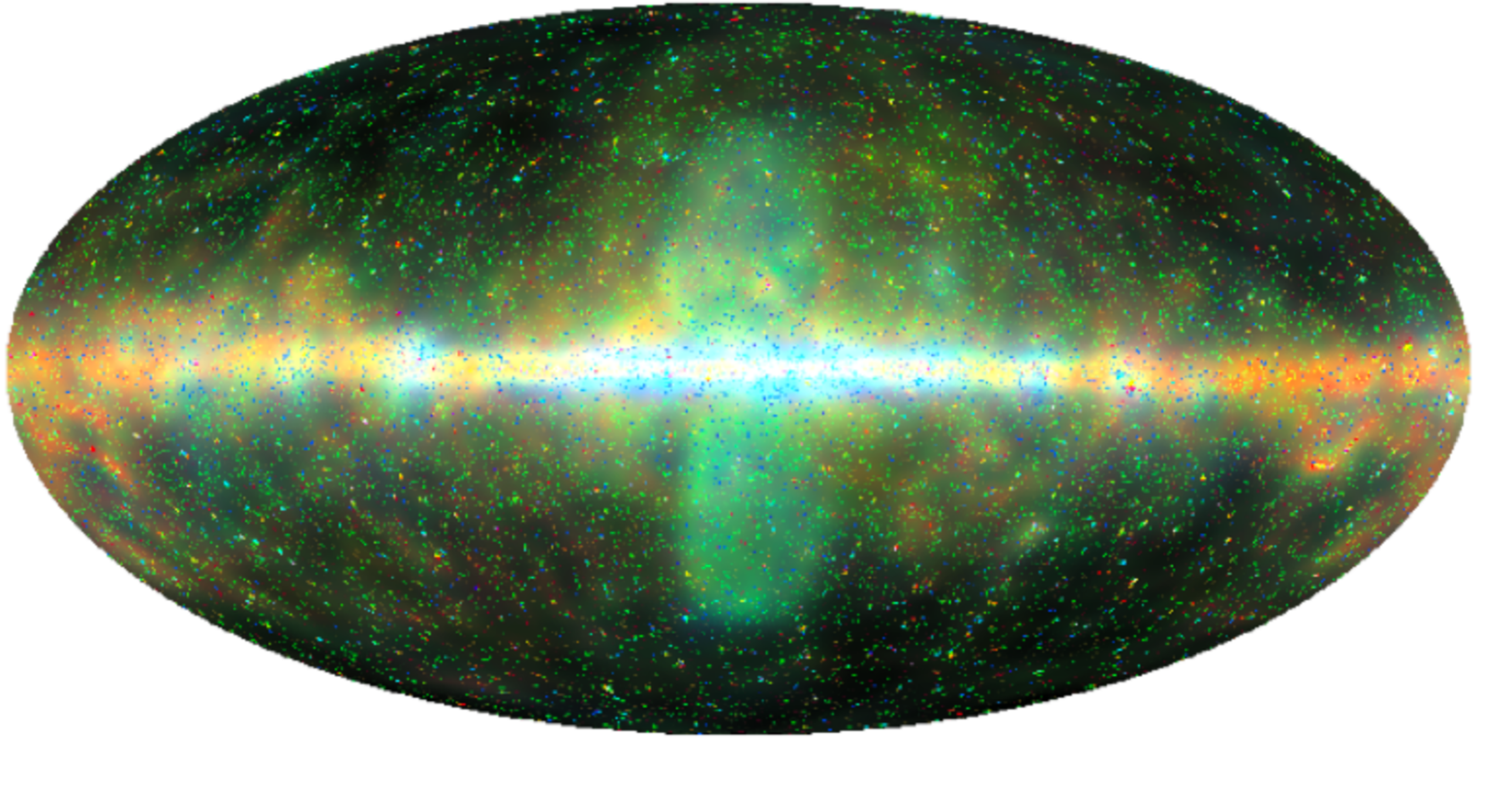}}
\resizebox{0.52\textwidth}{!}{
\includegraphics{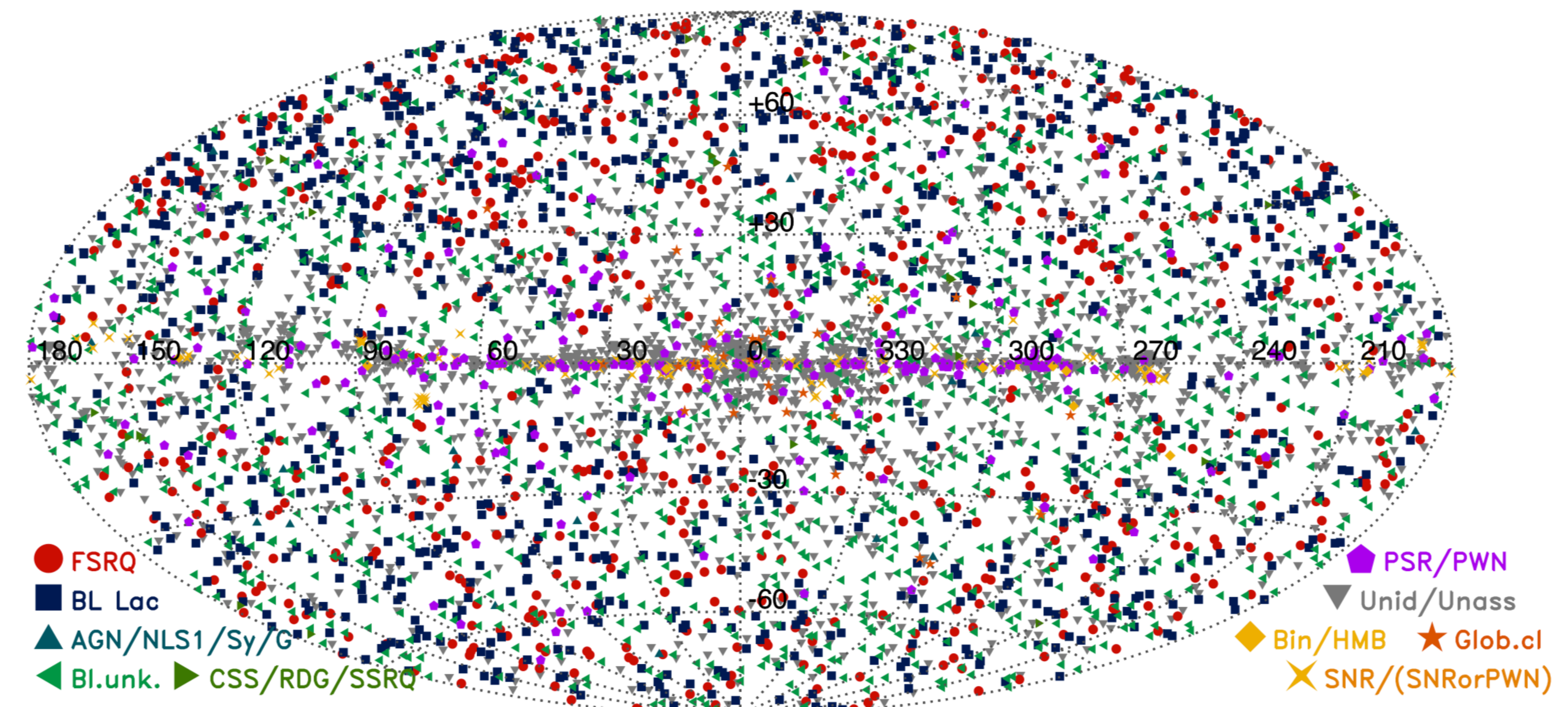}}
\caption{Left panel: Gamma-ray sky above 600 MeV in Galactic coordinates \cite{buhler}. Both the diffuse Galactic emission and point sources are visible. Intensity indicates the (logarithmic) brightness of the flux, red corresponds to low-energy gamma rays around 1 GeV, and blue to gamma rays up to 300 GeV. Right panel: sources detected by \fermilat  above 100 MeV in 8 years.}
\label{gamma_sky}       
\end{figure}
\subsection{The diffuse emission}
Below TeV energies, the dominant contribution (about 80\% in the GeV range) to the gamma-ray flux is due the Galactic diffuse emission, observed for the first time by OSO 3, and then confirmed also by SAS-2 and COS-B. Fig.~\ref{gamma_sky} (left panel) shows this diffuse emission as measured by \fermilat. 
Above 30 MeV, a large fraction of this emission is presumably due to: (1) CRs  interacting with the Galactic interstellar gas, via neutral pion decay, and contributing to the soft component of the gamma emission; (2) CR electrons scattering off interstellar radiation fields, via IC and contributing to the hard component of the gamma emission. This Galactic diffuse emission has been observed also at TeV energies by H.E.S.S. in the innermost part of the Galaxy~\cite{hesssurvey} and also by HAWC in a large scale region on the Galactic Plane~\cite{hawcsurvey}.

In addition to the diffuse emission, two huge bubble-like structures, extending $\sim50^\circ$ above and below the GC, have been observed~\cite{fbobs1,fb1,fbobs2}.
The gamma-ray emission from these structures, dubbed the \textit{Fermi Bubbles}, exhibits a power-law spectrum with spectral index $\Gamma=1.9\pm0.2$, significantly harder than the spectrum of the diffuse emission from the Galactic disk, and a cut-off energy of $(110\pm50)$~GeV~\cite{ack_fermib}. These structures are visible in the left panel of Fig.~\ref{gamma_sky}. A possible origin of the bubbles includes CR acceleration by the SMBH Sgr A*~\cite{fb1} at the center of our Galaxy, an era of starburst activity, or an accumulation of CR for a long time from the regular star formation near the GC~\cite{fb2}. 

A (mostly extragalactic)  gamma-ray background, briefly  called  IGRB (isotropic gamma-ray background), was detected for the first time by SAS-2; \fermilat has then performed a measurement between 100~MeV and 820~GeV~\cite{Ackapj799}, showing that the IGRB is characterized by a power-law spectrum, with spectral index $\Gamma=2.32\pm0.02$ and an exponential cut-off at $279\pm52$~GeV. This cut-off could be explained by a single dominant extragalactic population with EBL attenuation. In general, the IGRB is composed by unresolved extragalactic emissions,   and by residual Galactic foregrounds. The total intensity attributed to the IGRB is $(7.2\pm0.6)\times10^{-6}$ cm$^-2$ s$^-1$ sr$^{-1}$ above 100 MeV.
\subsection{Steady sources and transient events}
\begin{figure}
\centering    
\resizebox{0.46\textwidth}{!}{
\includegraphics{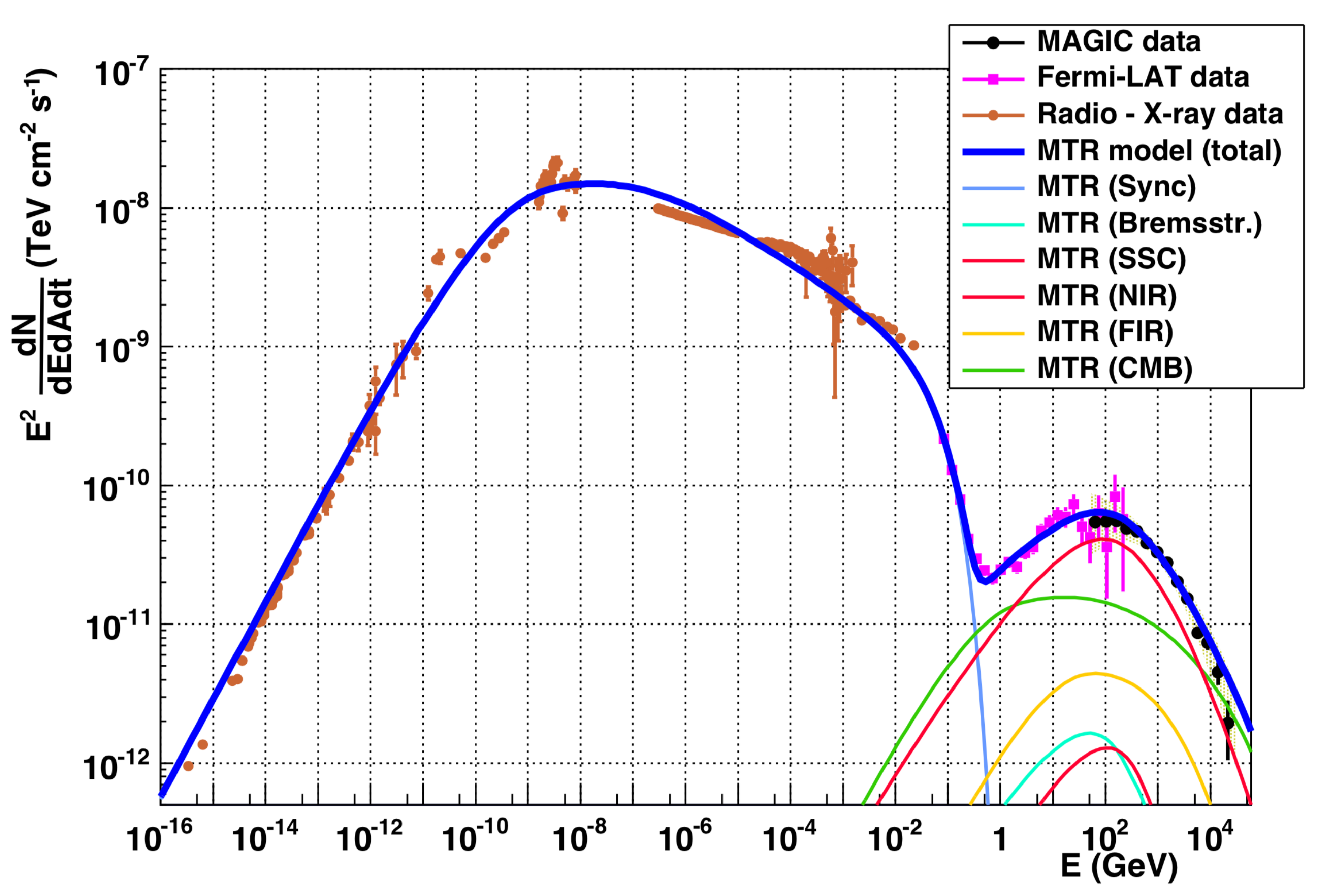}}
\resizebox{0.4\textwidth}{!}{
\includegraphics{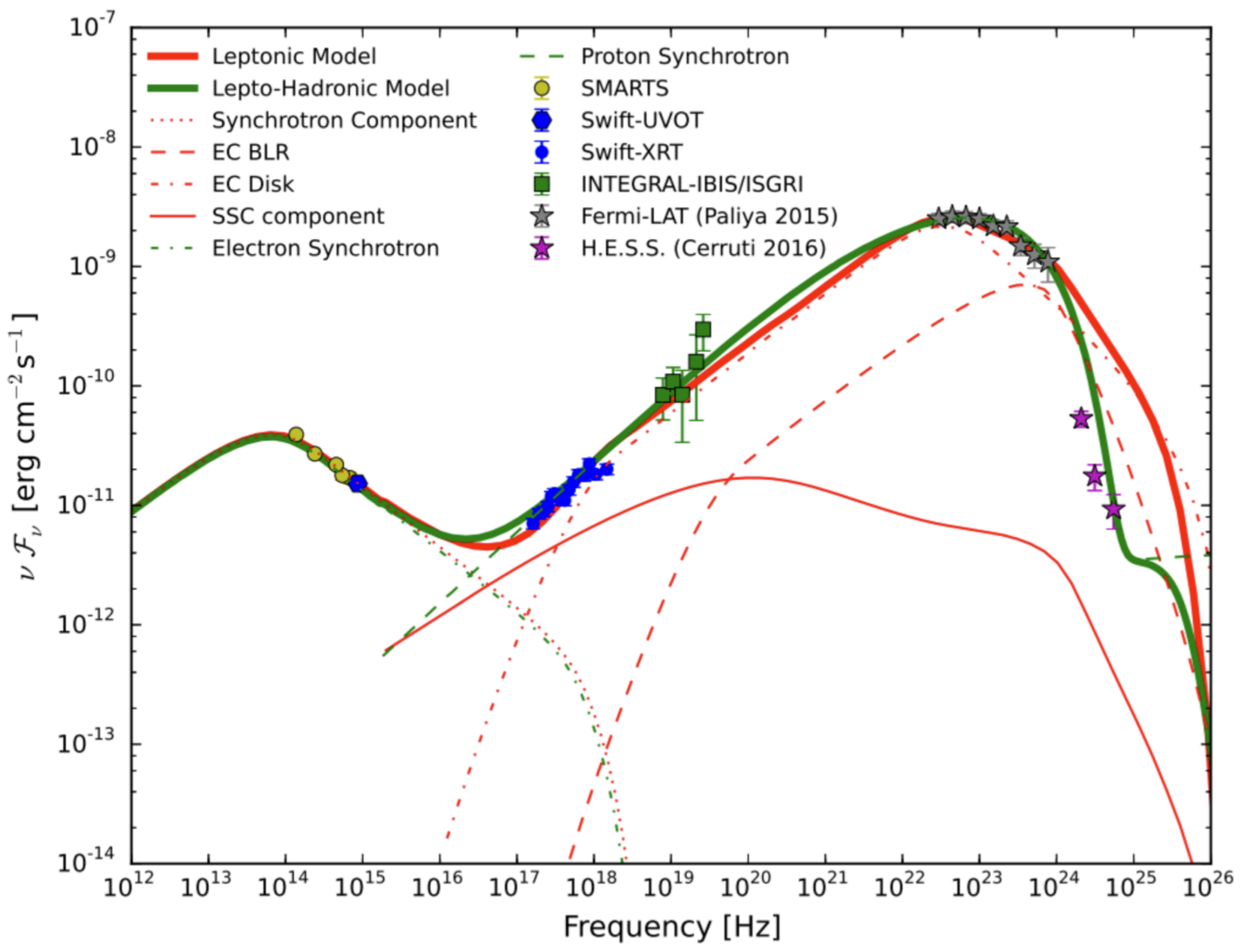}}
\caption{Two typical sources respectively Galactic and extragalactic. Left panel:  multi-wavelength SED of the Crab Nebula, from radio to gamma rays \cite{magic_crab}. Right panel: Quasi-simultaneous SED of the blazar 3C 279, along with the leptonic (red solid) model and the lepto-hadronic (green solid) model, as presented in \cite{bottacini}.}
\label{sources}       
\end{figure}

One of the most impressive results about the gamma-ray sky regards the plethora of individual emitters that have been discovered over the last decade. More than $\sim$ 5000 sources above 100 MeV have been identified up to now thanks to the \fermilat 8-year observations (Fig.~\ref{gamma_sky}, right), which will converge soon to the Fermi 4th catalog. The third \fermilat catalog~\cite{3fgl}, corresponding to 4 years of data taking, contained more than 3000 sources. About $\sim$ 80\% are associated to extragalactic objects, and a significative fraction of Galactic objects is associated to pulsars. It is worth mentioning the most studied Galactic source of gamma rays: the \textit{Crab Nebula}. It is a typical PWN,  the leftover of the supernova explosion that occurred in 1054 A.D., and it is powered by the pulsar PSR B0531+21 at its center. Its multi-wavelength SED is shown in Fig.~ \ref{sources} (left panel)~\cite{magic_crab}.

At higher energies ($\gtrsim$~100~GeV), $\sim$~200 sources have been detected and are collected in the TeVCat webpage~\cite{tevcat}. About half are objects in our Galaxy, including SNRs, pulsars, PWNe, binaries and a significative fraction of unidentified sources. The remaining half is of extragalactic origin, but the angular resolution of the current detectors (slightly better than $0.1^\circ$) is not good enough to associate them with particular points in the host galaxies. These gamma-ray extragalactic sources are starburst galaxies and  AGN ($\sim$ 37\%). According to the generally accepted hypothesis, a SMBH, having up to $\sim10^9$ solar masses ($M_\odot$), resides at the core of an AGN. As material falls into the SMBH,  gravitational energy is released and some of the energy can be converted into kinetic energy of an outflow, forming well-collimated jets of plasma with relativistic speed. The observed extragalactic TeV sources are mostly blazars, a particular class of AGN whose jet points toward the observer. Fig.~\ref{sources} (right panel) shows the multi-wavelength SED of the blazar 3C 279, as presented in \cite{bottacini}. In particular, it is a \textit{flat spectrum radio quasar} (FSRQ), a sub-class of blazars characterized by strong broad emission lines in the optical spectrum. 3C 279 was the first (of currently only 7) FSRQs detected by ground-based IACTs and it is one of the ideal target for multi-wavelength studies.

\begin{figure}
\centering    
\resizebox{0.55\textwidth}{!}{
\includegraphics{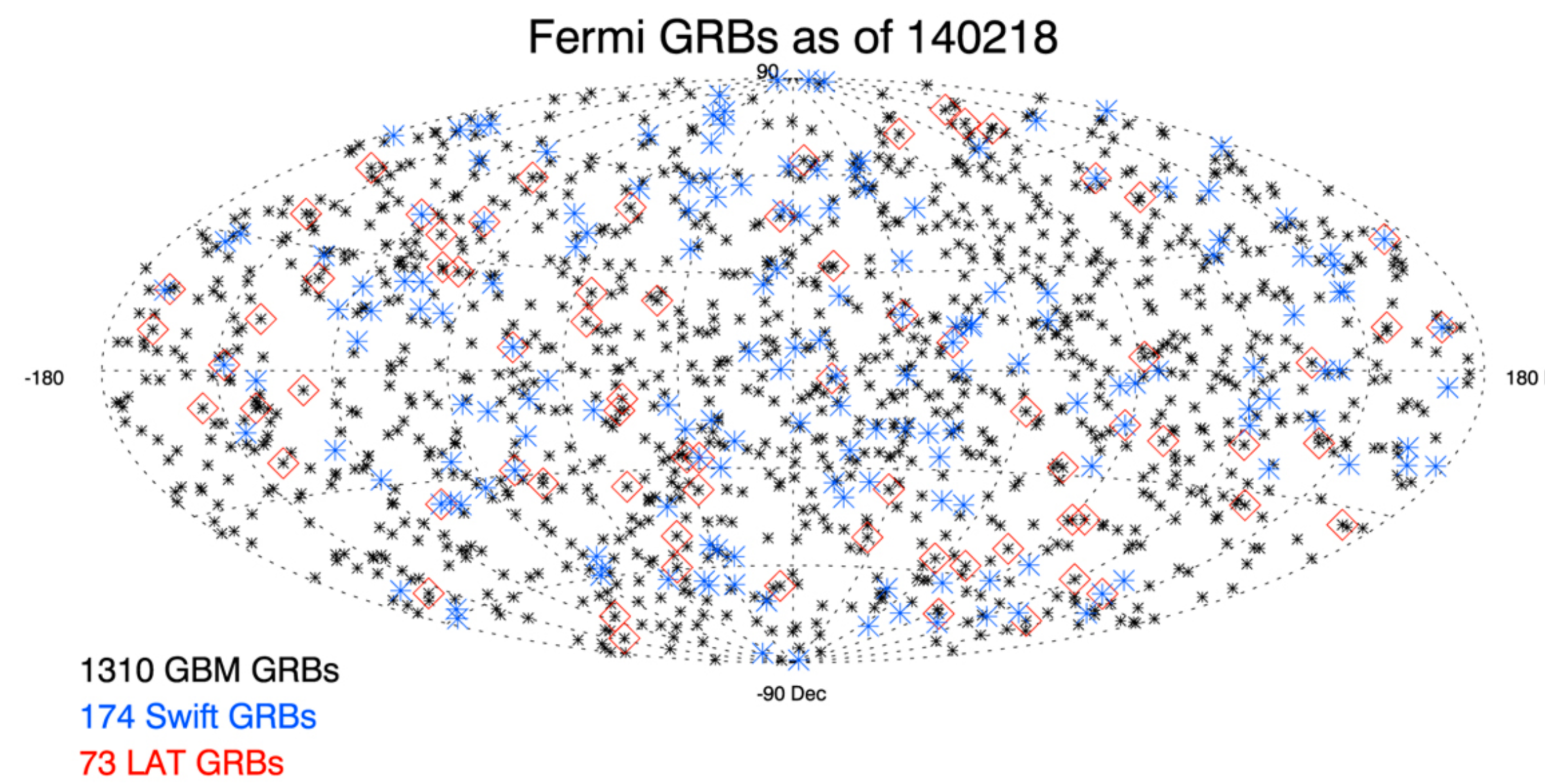}}
\caption{Skymap of GRBs detected by \fermi (both GBM and \fermilat). Some events recorded also by the Swift satellite are shown. Credit: NASA.}
\label{grb}       
\end{figure}

Another type of extragalactic objects contributes to the gamma-ray sky: the so-called \textit{gamma-ray bursts} (GRBs). These are transient events, recorded almost daily,  lasting from fractions of a second (the so-called ``short'' GRBs, recently associated to neutron star-neutron star mergers \cite{GW170817}), to a few seconds and
more (``long'' GRBs), associated to the collapse of a very large mass star ($\sim 10^2$ $M_\odot$), following a very energetic supernova (a ``hypernova'').
They are  often followed by ``afterglows'' after minutes, hours, or days.  The energy spectrum is non-thermal and varies from event to event, peaking at around a few hundred keV and extending up to several GeV. A few of them per year have energy fluxes and energies large enough that the \fermilat can detect them. Fig.~\ref{grb} shows a map of the GRBs detected up to now by both detectors onboard \fermi and by Swift.
\section{Multi-messenger astronomy}\label{sec:multi}
\subsection{Links with cosmic rays}\label{sec:CR}
Among the candidate CR accelerators, several sources have been studied in order to find the relation between gamma rays and charged particles. In the Milky Way in particular, gamma-ray emission from pulsars and PWNe is thought to originate from CR electrons. Leptonic models therefore usually explain the observed SED. SNRs are instead, since the hypothesis formulated by Baade and Zwicky in 1934, thought to be accelerators of CR ions. According to the relation in Eq.~\ref{Hillas}, these objects could accelerate CRs up to PeV energies and therefore they are candidate \textit{``PeVatrons''}. This hypothesis has a twofold justification. From one side, SNRs are natural places in which strong shocks develop with high speed $\beta$.  They can  account for the required energetics and for the energy density of CRs. In addition, the SNR environment offers the targets, like molecular clouds and photon fields, for the interaction with CRs.
Two classes of SNRs have been identified. The first one is mainly detected at GeV energies. They are middle-aged SNRs (t$_{\mathrm{age}}>$~10$^4$~yrs), characterized by the presence of molecular clouds. Objects like IC 443, W44 and W51C are the brightest objects belonging to this category. The second kind of SNRs has harder spectra and is often observed also at TeV energies. Examples of this type are Tycho's SNR, Cas A, SN 1006.

\begin{figure}
\centering    
\resizebox{0.7\textwidth}{!}{
\includegraphics{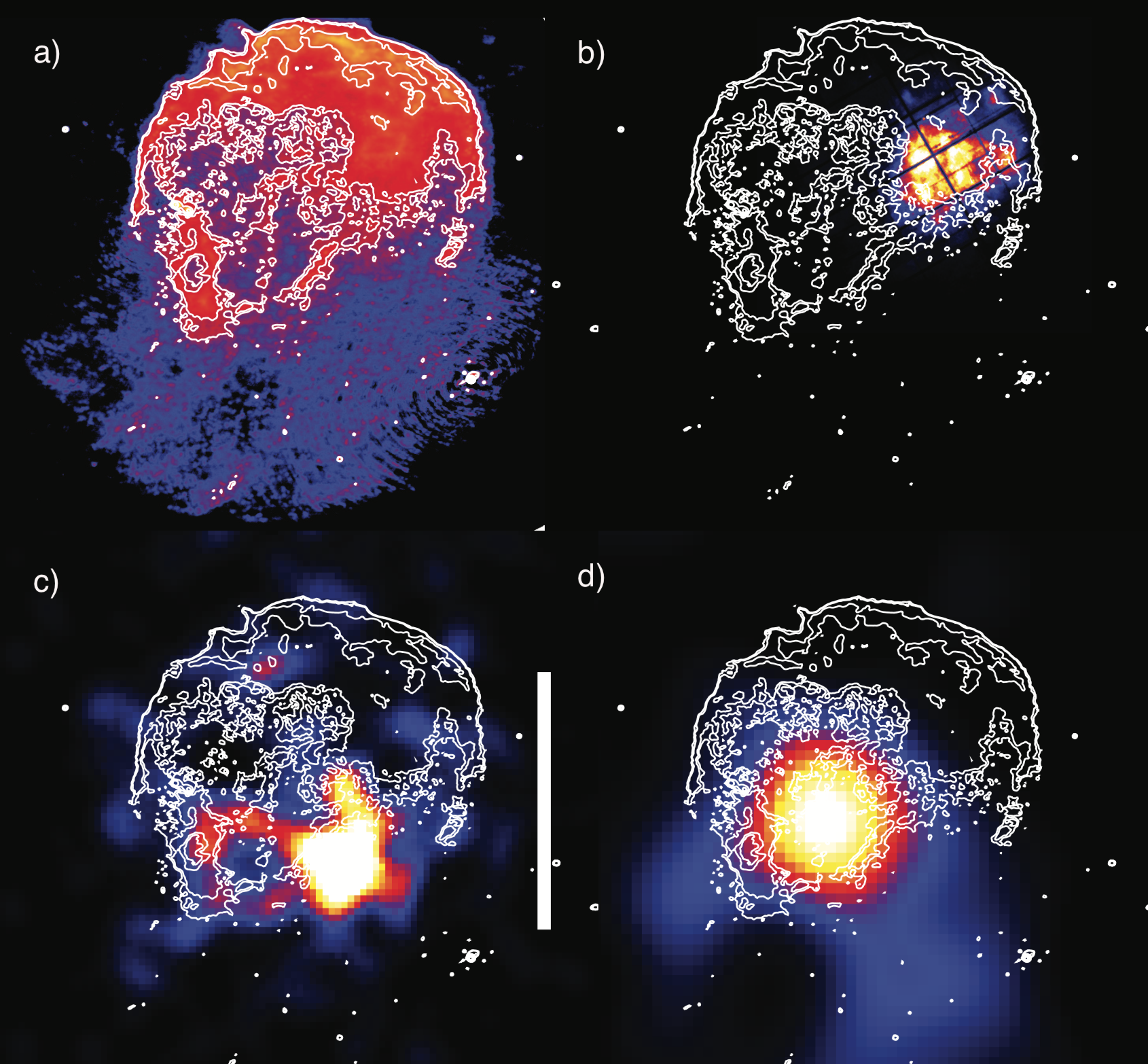}
}
\caption{Multi-wavelenght data of IC443 in Galactic coordinates~\cite{funk}. a) VLA 330 MHz continuum emission; b) XMM observations (0.3-8 keV); c) \fermilat observations ($>$ 10 GeV) d) MAGIC observations above 380 GeV}
\label{fig:IC443}       
\end{figure}

For IC 443 and W44, thanks to Agile and \fermilat, indications of the  pion bump, discussed in Sect.~\ref{sec:had}, and an appropriate morphology with emission positionally consistent with molecular clouds,  have been found. Indeed current IACTs can resolve the morphology of some SNRs, allowing the study of the medium surrounding the source and the acceleration. Fig.~\ref{fig:IC443} shows an example of this kind of study, where IC 443 SNR is shown in Galactic coordinates, for different energy bands, going from radio to very high energy~\cite{funk}. All this evidence  proves the hadronic origin of part of the gamma rays, although it is still unclear what fraction~\cite{bookap}. The spectrum of these classes of sources generally exhibits  a cut-off in the GeV energy range, therefore well below the PeV region of the CR spectrum.

Young SNRs could be a more suitable site of CR acceleration, given their very high energy emission. However, for the whole sample of young SNRs detected up to now the maximum energy observed in gamma rays is much lower than 100 TeV~\cite{park}.

As a general remark, we can state that there is no doubt that SNR accelerate (part of the) Galactic CR, the open questions being: which kind of SNR; in which phase of its evolution a SNR really does accelerate particles; and if the maximum energy of these accelerated particles can go beyond $\sim$1 PeV. Answers to these questions should provide insight on the nature, the energy and the composition of the CR spectrum at the knee.

\begin{figure}[h]
\resizebox{0.70\textwidth}{!}{
\includegraphics{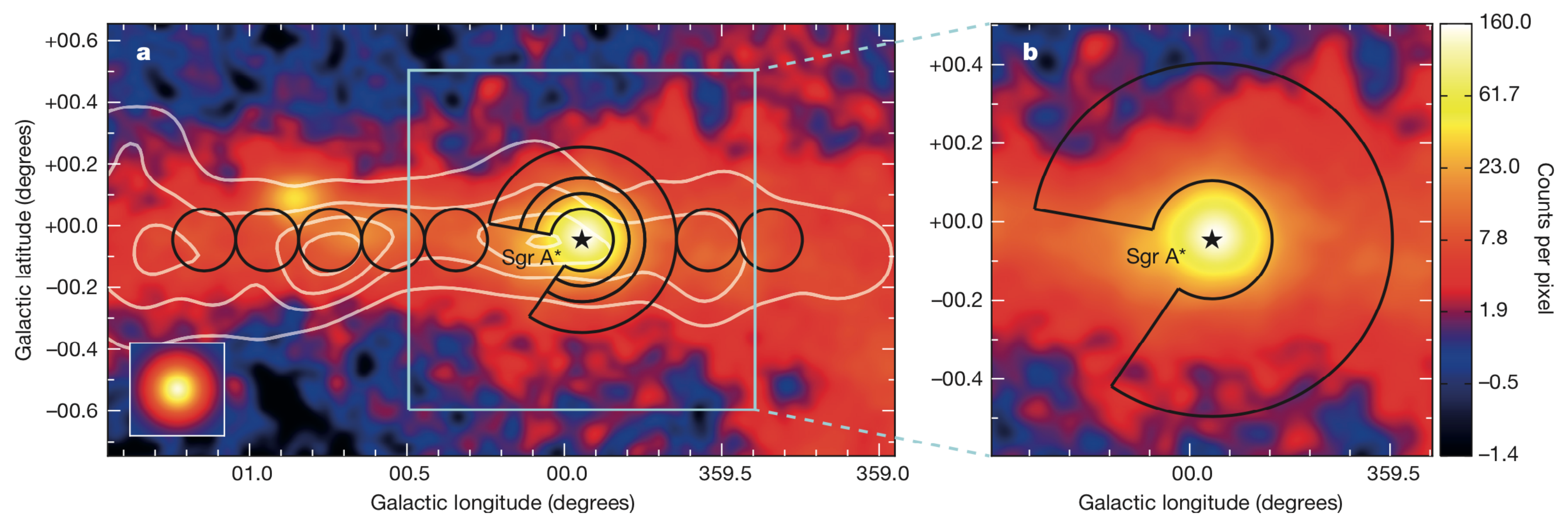}}
\resizebox{0.25\textwidth}{!}{
\includegraphics{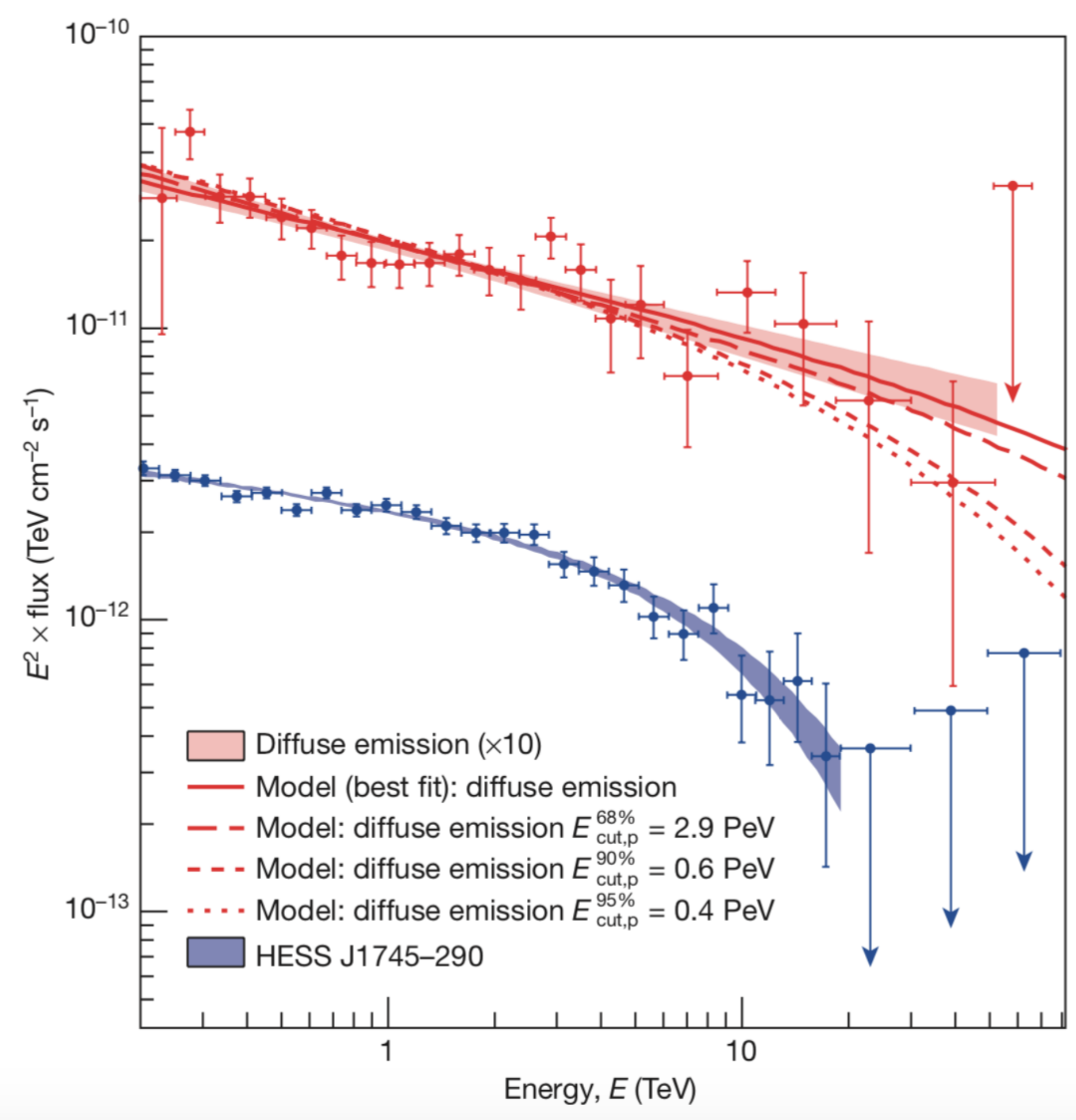}}
\caption{Left panel: Very-high energy gamma-ray image of the GC region. The black lines show the regions used to calculate the CR energy density throughout the central molecular zone. White contour lines indicate the density distribution of molecular gas. The inset shows the simulation of a point-like source. The inner $\sim 70$~pc and the contour of the region used to extract the spectrum of the diffuse emission are zoomed. Right panel: Very-high energy gamma-ray spectra of the diffuse emission and of the source HESS J1745-290, positionally consistent with the GC. The Y axis shows fluxes multiplied by a factor $E^2$,  in units of TeV~cm$^{-2}$s$^{-1}$.  Arrows represent 95\% C.L. flux upper limits.  The red lines show the numerical computations assuming that gamma rays result from the decay of neutral pions produced by proton-proton interactions. The fluxes of the diffuse emission spectrum and models are multiplied by 10~\cite{abram}.}\label{fig:peva1}
\end{figure}

In this context, we mention the interesting result reported by H.E.S.S. about the study of the GC \cite{abram}. It has been shown that the gamma-ray emission is compatible with a steady source accelerating CRs up to PeV energies within the central 10~pc of the Galaxy.  The SMBH Sgr A* could be linked to this possible PeVatron. 
The energy spectrum of  the diffuse gamma-ray emission (Fig.~\ref{fig:peva1}, right) has been extracted from an annulus centered at Sgr~A* (see Fig.~\ref{fig:peva1}, left). The best fit to the data is found for a power-law spectrum with spectral index  $\Gamma\simeq$2.3, extending up to energies of tens of TeV, without any indication of a cut-off.  Such hard power-law spectrum can imply that the spectrum of the parent protons extends to energies close to 1 PeV.  
Although its current rate of particle acceleration is not sufficient to provide a substantial contribution to Galactic CR, Sgr A* could have plausibly been more active over the last $\gtrsim 10^{6-7}$~years. However, this hypothesis is speculative; moreover, the identification of the source remains unclear, since the GC region is not well resolved.

The relation  in Eq.~\ref{Hillas} indicates that AGN are plausible candidates for CR acceleration up to the ultra-high energies. Although the spatial resolution of gamma-ray telescopes is not yet good enough to study the morphology of extragalactic emitters, a recent study of a flare from the nearby galaxy M87 (at a distance of about 50 Mly) by the main gamma telescopes plus the VLBA radio array has shown, based on the VLBA imaging power, that this AGN accelerates particles to very high energies in the immediate vicinity  (less than 60 Schwarzschild radii) of its central black hole (BH). This galaxy is very active: its BH, of a mass of approximately $7\times10^9$~$M_\odot$, has an accretion rate of~$\sim$~2-3~$M_\odot$/year. A jet of energetic plasma originates at the core and extends outward at least 5000 light-years~\cite{m87cr}.

The lack of simultaneous observations of neutrinos and GRBs~\cite{halzen} allows to set a stringent limit (about 6\%) to the maximum fraction of CRs above the knee coming from GRBs.

\subsection{Links with astrophysical neutrinos}\label{sec:neutrino}
As explained in Sect.~\ref{sec:had}, in a bottom-up scenario, astrophysical neutrinos of TeV-PeV energies (and possibly beyond) are expected to be produced by astrophysical sources together with gamma rays, via hadronic processes. 
Neutrinos are not deflected by magnetic fields, nor they are absorbed up to energies of $\sim10^{16}$~eV, because of their small interaction cross-section. On the other hand, compared to photons, they are extremely difficult to detect, being their cross-section the lowest among elementary particles. 

High-energy astrophysical neutrinos were detected only in 2013, thanks to the IceCube experiment, located in the South Pole, that collects $\simeq$~1~event/month of this kind. In particular, the neutrino flux has been observed from about 10 TeV to several PeV, but its origin is still unknown: no sources have been identified~\cite{halzen,IceCube}. 

A recent IceCube analysis suggests that blazars contribute at most 27\% of the observed IceCube intensity~\cite{ICbl}.
However, neutrinos could be emitted during transient events like flares. In order to probe this scenario,  simultaneous observations of neutrino and gamma-ray signals are fundamental. 
A potentially compelling evidence has been found on September 2017, when \fermilat~\cite{fermi_nu} and MAGIC~\cite{magic_nu} detected with a significance larger than 5$\sigma$ an enhanced gamma-ray emission from the source TXS 0506+056 at a distance $z\sim0.34$, positionally consistent with the $\sim$ 300~TeV neutrino IC170922A~\cite{GCN} detected by IceCube. 

 This result highlights the importance of performing simultaneous observations of both gamma rays and neutrinos, in order to show the hadronic emission process at work in astrophysical sources.  
\subsection{Links with gravitational waves}\label{sec:GW}
Gravitational waves were predicted by  Einstein's theory of the General Relativity. About one century later, the direct evidence of their existence has been firmly established by the LIGO/Virgo collaboration, opening a new and exciting era for astrophysics~\cite{GW150914}. 
On September 14th, 2015, the two LIGO detectors observed simultaneously a large and clear gravitational wave signal (labelled as GW150914) that matches the theoretical prediction for the coalescence of a binary system of BHs, with $\sim$~30~$M_\odot$ each.  After this discovery (that earned Rainer Weiss, Barry C. Barish  and Kip S. Thorne the Nobel Prize in Physics 2017), four more events were detected, again interpreted as the coalescence of a binary system, with BH masses around 15-30~$M_\odot$. 

A special event for multi-messenger astronomy was recorded on August 17th, 2017, called GW170817. It is the first observation of a single astrophysical source through both gravitational and electromagnetic waves~\cite{GW170817}.

\begin{figure}
\centering    
\resizebox{0.90\textwidth}{!}{
\includegraphics{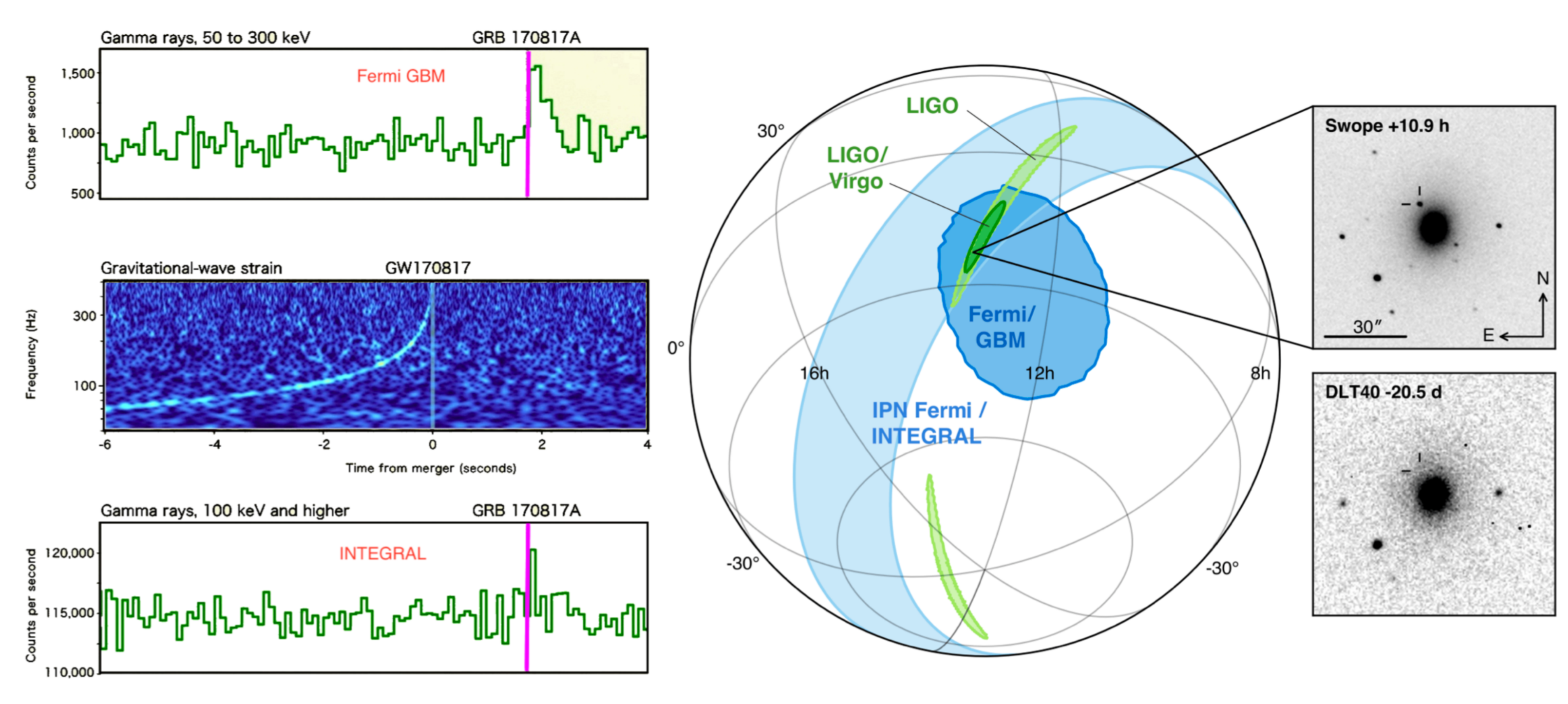}
}
\caption{The signals detected by  \fermi GBM (top left), by LIGO/Virgo (center left), and by \integral (bottom left); the 90\% location contour regions of the GW170817 / GRB 170817A / SSS17a event as determined by LIGO, LIGO-Virgo, 
INTEGRAL, \fermi. The insets show the location of the NGC 4993 galaxy in the images of the Swope (top right) and of the DLT40 (bottom right) optical telescopes respectively 10.9 hr after and 20.5 days before the GW observation. The perpendicular lines indicate the location of the transient in both images. Courtesy S. Ciprini, ASI.}\label{fig:firstmw}
\end{figure}

LIGO/Virgo detected a gravitational wave signal possibly associated with the merger of two neutron stars (NS) and $(1.75\pm0.05)$~s later  \fermi GBM  and \integral SPI/ACS observed independently in the same sky region (Fig.~\ref{fig:firstmw}) a short ($\sim$2~s) GRB (GRB 170817A). The time-averaged spectrum of GRB 170817A is well fitted by a power law function, with an exponential cut-off at $\sim 80$ keV. The masses of the initial NS were estimated to be in the range [1.36, 2.26] $M_\odot$ and [0.86, 1.36] $M_\odot$ respectively, while the final  mass was estimated to be $2.82^{+0.47}_{-0.09} M_\odot$.  These observations were followed by an extensive multi-messenger campaign covering all the electromagnetic spectrum, as well as the neutrino channel: a bright optical transient (SSS17a) was discovered by the Swope Telescope in South America and shortly after by  five more teams. The latter is associated with the NGC 4993 galaxy, located at a distance of 40 Mpc. The follow-up was then done by ground and space observatories all around the world: X-ray and radio counterparts were discovered respectively $\sim$~9~days and $\sim$16~days after the merger, while  no neutrino  candidates were seen. 

The NS-NS merger event is thought to result in a ``kilonova'', characterized by a short GRB followed by a longer optical afterglow; the event has been observed at an angle $\sim 30^\circ$ from the  axis of the jetted emission, which allows extrapolating the on-jet energy cut-off to a few MeV~\cite{grenot}. This fact enforces the scientific case for a MeV-GeV gamma-ray experiment \cite{wb}. 
A total of 16~000 times the mass of the Earth in heavy elements is believed to have formed; for some of them spectroscopical signatures have been observed.

The scientific importance of this event is huge, with implications for gamma-ray astronomy (it provides a strong evidence that mergers of binary stars are the cause of short GRBs). Unlike all previous detections, corresponding to BH-BH merging and not expected to produce a detectable electromagnetic signal, the aftermath of this merger was seen by 70 observatories   across the electromagnetic spectrum, marking a significant breakthrough for multi-messenger astronomy and opening a new era. Several events of this kind can be expected in the future.

In addition,  this joint multi-messenger observation provides a limit on the difference between the speed of light and that of gravity.  The relative difference between the speeds of gravitational and electromagnetic waves, $|v_{GW} - v_{EM}|/c$, is constrained to be smaller than $\sim 10^{-15}$.

An electromagnetic emission of the order of a few tens MeV can be expected in the case of the merging of a NS with a BH of $\sim 14-20 M_\odot$ \cite{nsbh}.

\section{The search for Dark Matter}\label{sec:dm}

\begin{figure}
\centering
\resizebox{0.4\textwidth}{!}{
\includegraphics{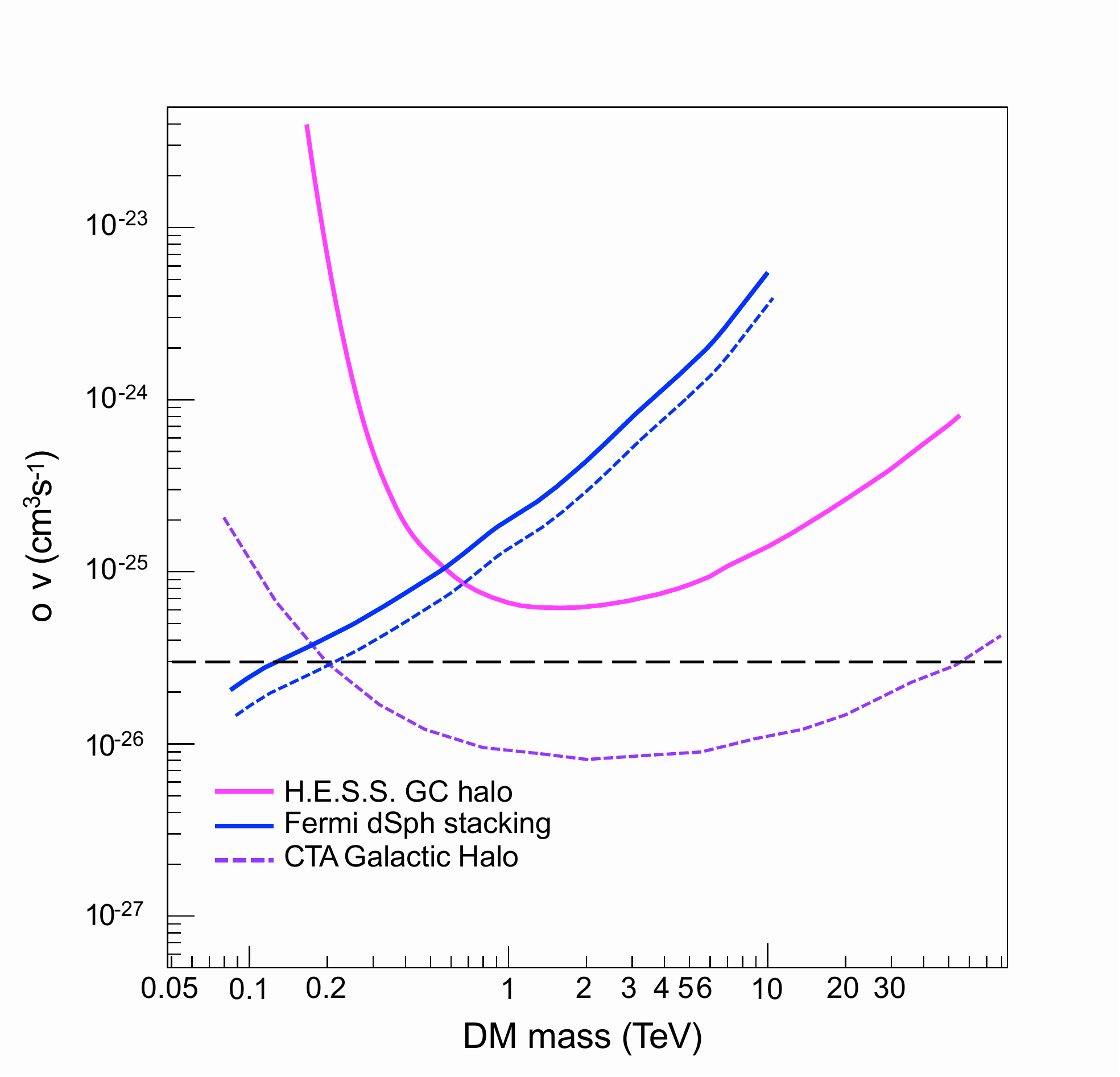}
}
\caption{Sensitivities to DM in terms of $< \sigma v >$ from the observation of the  Galactic halo (present results from H.E.S.S., continuous line,  and expected results from 3 years of operation of CTA South, dotted line), and from a stacked sample of dSPHs ($Fermi$-LAT). The LAT lines are relative to 6 years of data analysis (continuous, upper line; this is the present result) and to an extrapolation to 10 years of  analyzed data (dotted, lower).
The horizontal dashed line
indicates the thermal velocity-averaged cross-section. From \cite{CTA2} and  $Fermi$-LAT publications.}\label{fig:dmlim}
\end{figure}

DM densities are unknown in the innermost regions of galaxies, where most of the signal should come from: data allow only the computation
in the halos, and models helping in the extrapolation to the centers frequently disagree. Observations of galaxy rotation curves favor constant density
cores in the halos; unresolved ``cusp''
substructures can have a very large impact, but their existence is speculative -- however, since they exist for baryonic matter, they are
also likely  for DM. This uncertainty is typically expressed by the so-called ``boost factor,'' defined as the ratio of the true, unknown,
line-of-sight integral to the one obtained when assuming a smooth
component without substructure.
As a consequence of all uncertainties described above, the choice of targets is somehow related to guesses, driven by the knowledge of locations where one expects large ratios of gravitating to luminous mass. The main  targets are:
\begin{itemize}
\item{\em Galactic center.}
The GC is expected to be the brightest source of DM annihilation. However, the many astrophysical sources of gamma rays
in that region complicate the identification of DM. In the GeV region the situation is further
complicated by the presence of a highly structured and extremely
bright diffuse gamma-ray background arising from the interaction of
the pool of CRs with dense molecular material in the inner
Galaxy. 
To limit problems, searches for DM
annihilation/decay are usually performed in regions 0.3$^\circ$--1$^\circ$ away form the central BH.

At TeV energies, IACTs detected a point source compatible with the position of  the SMBH at the center
of our Galaxy  and a diffuse emission coinciding with molecular material in the Galactic
ridge. The GC source has a featureless power-law spectrum at TeV energies with an exponential cut-off at
 $\sim$$10$ TeV not indicating a DM scenario; the signal is usually attributed
  to the SMBH Sgr~A$^{\star}$  or to a PWN in that region. Searches have been performed
for a signal from the Galactic DM halo  close to the core; no signal has been found.
There have been several claims of a signal in the GC region. An extended
signal coinciding with the center of our Galaxy, corresponding to a WIMP of mass about 40 GeV/$c^2$ was reported
above the Galactic diffuse emission---however, the interaction of freshly produced CRs with interstellar
material is a likely explanation.
\item{\em Dwarf Spheroidal Galaxies.} Dwarf spheroidal galaxies (dSph) are a  clean environment to search
for DM annihilation:   astrophysical backgrounds that
produce gamma rays are expected to be negligible. The DM content can  be determined from stellar dynamics and
these objects have been found to be the ones with the largest
mass-to-light ratios in the Universe, and uncertainties on the boost factor are within one order of magnitude.
Some three-four dozens of  dwarf satellite galaxies of the Milky Way are currently known
and they are observed both by ground- and space-based gamma detectors. No  signal has been found, and stringent limits have been calculated. In particular, a
combined (``stacked'') analysis of all known dSphs with the $Fermi$-LAT satellite
has allowed a limit to be set
below the canonical thermal relic production cross section of $3\times
10^{-26} \mathrm{cm}^3 \mathrm{s}^{-1}$ for a range of WIMP masses
(around 10~GeV) in the case of the annihilation into
${b}\bar{{b}}$.
\item{\em Galaxy clusters.} Galaxy clusters are groups  of   hundreds to thousands  galaxies
bound  by gravity. Nearby clusters  (10--100 Mpc) include
the Virgo, Fornax, Hercules, and  Coma clusters. 

Galaxy clusters are much more distant than dSphs or any of the other targets   generally used for DM searches with gamma rays; however, like dSphs, astrophysical dynamics show
 that they are likely to be DM dominated---and if DM exists, one of the largest accumulators. The range of likely boost
factors due to unresolved DM substructures can be
large; however, when making conservative
assumptions, the sensitivity to DM is several orders of magnitude away from the canonical thermal relic
interaction rate.
\item{\em Line Searches.} The annihilation of WIMP pairs into $\gamma X$ would lead to monochromatic
gamma rays with $E_\gamma = m_\chi(1-m_X^2/4m_\chi^2)$ (with $m_\chi$ WIMP mass). Such a signal
would provide a smoking gun  since
astrophysical sources could  hardly produce it, in particular if  found in several
locations. 
\end{itemize}
A summary of the results on WIMPs from gamma-ray experiments is shown in Fig.~\ref{fig:dmlim}.

\section{Future directions}\label{sec:future}
 The current generation of detectors, both in space and on ground, is providing a wealth of information about the non-thermal Universe in the GeV-TeV energy band, as highlighted in the previous sections. However  a key piece is missing: the MeV energy range. This  scarcely explored domain is important since it is characteristic  of nuclear transitions and of the nuclear de-excitation of molecular clouds excited by colliding CRs; in addition, it is the energy range where one expects the exhaustion of the electromagnetic counterpart of gravitational wave events and where one expects gamma rays from the conversion of axions in the core of supernovae.
 The MeV band is therefore full of scientific information \cite{wb}.
 Currently, the germanium detector (SPI) onboard \integral covers the 18 keV to 8 MeV range, but without imaging capabilities.   COMPTEL onboard CGRO was the last instruments to perform measurements of the gamma-ray sky in the MeV energy range. The sensitivities of both SPI and COMPTEL are shown in Fig.~\ref{sensitivity}, together with that of the e-ASTROGAM mission proposed by a mostly European consortium \cite{ea}.  e-ASTROGAM is designed to have a large FoV ($>$2.5 sr) and broad energy coverage (0.3 MeV to 3 GeV), with one-two orders of magnitude improvement in continuum sensitivity in the range 0.3 MeV - 100 MeV compared to previous instruments. Such a kind of detector would therefore combine Compton and pair-production detection techniques. AMEGO is a similar project for MeV astrophysics, proposed for consideration in NASA's 2020 decadal review \cite{amego}. 
 
At the multi-GeV scale, the  
Chinese-Italian space mission HERD \cite{HERD}, scheduled for launch in 2025, will improve some of the aspects of $Fermi$, e.g. calorimetry. For sure a satellite in the GeV region with sensitivity comparable with $Fermi$ will be needed in space ($Fermi$ could in principle operate till 2028).
 
 For what concerns  ground-based observatories at the TeV scale, the Cherenkov Telescope Array (CTA)\cite{CTA1,CTA2}  represents the future for  IACTs, with
  an order of magnitude improvement in sensitivity over existing instruments.
An array of tens of telescopes will detect gamma-ray-induced showers over a large area on the ground,
increasing the efficiency and the sensitivity, while providing multiple views of each cascade. This will result in
both improved angular resolution and better suppression of CR background events. In particular, three types of telescopes are foreseen and listed in the following.
\begin{itemize}
\item The low energy (the goal is to detect showers starting from an energy of 20\,GeV)  instrumentation will consist of   23\,m large-size telescopes (LST) with a  FoV of about 4--5 degrees.
\item The medium energy range, from around 100\,GeV--1\,TeV, will be covered by medium-size  telescopes (MST) of the 12\,m class with a FoV of 6--8 degrees.
\item The high-energy instruments, dominating the performance above 10\,TeV, will be small size (SST, 4\,m in diameter) telescopes with a FoV of around 10 degrees.
\end{itemize}
CTA will be deployed in two sites. The Southern  site is less than 10 km southeast of the European Southern Observatory's (ESO's) existing Paranal Observatory in the Atacama Desert in Chile; it will cover about three square kilometers of land
with telescopes  that will monitor all the energy ranges in
the center of the Milky Way. It will consist of all three types of
telescopes with different mirror sizes (4 LSTs, 25 MSTs and 70 SSTs in the present design). The Northern hemisphere site is located on the existing   Roque de los Muchachos Observatory on the Canary island of La Palma,  close to MAGIC; only the two larger
telescope types (4 LSTs and 15 MSTs in the present design) would be deployed, on a surface of about one square kilometer.
These telescopes will be mostly targeted at extragalactic astronomy.
The telescopes of different sizes will be arranged in concentric
circles, the largest in the center. 
Different
modes of operation will be possible for CTA: deep field observation,
pointing mode, scanning mode -- also pointing to different targets.
 PeVatrons and the nature of the emitters in the Galaxy will be studied in detail. WIMPs will be tested with the ``right'' sensitivity up to 1 TeV.
CTA will be probably upgraded including state-of-the art photon detection devices of higher efficiencies with respect to the present ones; it can in principle operate till 2050, and it will work as an observatory.
 The sensitivity of CTA-South is shown in Fig. \ref{sensitivity} and compared to present detectors.
 
 At the multi-TeV to PeV scale, in the Northern hemisphere the LHAASO Observatory is in construction in China \cite{LHAASO}. LHAASO is a hybrid detector covering a total area of about $10^6$ m$^2$ with more than 5000 scintillation detectors, each of 1 m$^2$ area.
A central detector of 80~000 square meters (four times HAWC)  surface consists of water pools equipped with PMTs to study gamma-ray astronomy in the sub-TeV/TeV energy range.
About 1200 water tanks underground, with a total sensitive area of about 42~000 m$^2$, pick out muons, to separate gamma-ray initiated showers from hadronic showers. 
18 wide-FoV Cherenkov telescopes will complete the observatory.
 LHAASO will have the best sensitivity on gamma-ray initiated showers above some 10 TeV. One quarter of  the observatory should be ready by 2019, and  completion is expected in 2022. There is a strong case for a PeV wide-FoV detector in the Southern hemisphere in order to study the highest-energy emissions of accelerators in the Galaxy. Several collaborations (e.g. \cite{HAWCS,LATTES}) are proposing designs for such a detector, and convergence could be reached in the next years.

%

%
%

%
%

\end{document}